%% file: main.tex
\documentclass[journal]{IEEEtran}
\usepackage{amsmath,amsfonts,amsthm}
\usepackage{algorithmic}
\usepackage{algorithm}
\usepackage{array}
\usepackage{textcomp}
\usepackage{stfloats}
\usepackage{url}
\usepackage{verbatim}
\usepackage{graphicx}
\usepackage{cite}
\usepackage{xcolor}

\IEEEoverridecommandlockouts

\usepackage{xspace, subfigure, amssymb}
\usepackage[nolist,nohyperlinks]{acronym}
\usepackage{overpic}
\usepackage{lipsum}

\newcommand{\corollaryname}{Corollary}

\theoremstyle{plain}

\newtheorem{corollary}{Corollary}

\newcommand{\meter}{\,\mathrm{m}}
\newcommand{\km}{\,\mathrm{km}}
\newcommand{\mm}{\,\mathrm{mm}}
\newcommand{\dB}{\,\mathrm{dB}}
\newcommand{\dBm}{\,\mathrm{dBm}}

\newcommand{\MHz}{\,\mathrm{MHz}}
\newcommand{\GHz}{\,\mathrm{GHz}}

\begin{document}

\title{Spherical Wavefronts Improve MU-MIMO Spectral Efficiency When Using Electrically Large Arrays}

\author{Giacomo Bacci,~\IEEEmembership{Member,~IEEE,}
        Luca Sanguinetti,~\IEEEmembership{Senior Member,~IEEE} and  Emil Bj{\"o}rnson,~\IEEEmembership{Fellow,~IEEE}\vspace{-0.6cm}
        % <-this % stops a space
%\thanks{G. Bacci and L. Sanguinetti are with the Dip. Ingegneria dell'Informazione, University of Pisa, 56122 Pisa, Italy. E. Bj{\"o}rnson is with the Dept. Computer Science, KTH Royal Institute of Technology, Kista, Sweden.}% <-this % stops a space
%\thanks{Manuscript received Month xx, 2022.}
}

%% The paper headers
%\markboth{IEEE Wireless Communications Letters, Vol.~X, No.~Y, Month~2022}%
%{Bacci \MakeLowercase{\textit{et al.}}: Title}

%\IEEEpubid{0000--0000/00\$00.00~\copyright~2022 IEEE}

\maketitle

\begin{abstract}
%Wireless communication systems have almost exclusively operated in the far-field of antennas and antenna arrays, which is conventionally characterized by having propagation distances beyond the Fraunhofer distance. This is natural since the Fraunhofer distance is normally only a few wavelengths.
Modern \ac{MIMO} communication systems are almost exclusively designed under the assumption of locally plane wavefronts over antenna arrays. This is known as the \emph{far-field approximation} and is soundly justified at sub-6-GHz frequencies at most relevant transmission ranges. However, when higher frequencies and shorter transmission ranges are used, the wave curvature over the array is no longer negligible, and arrays operate in the so-called \emph{radiative near-field region}. 
%This is typically overlooked in the literature, though it may have a profound effect on the design and performance of current and future multiple antenna systems. 
This letter aims to show that the classical far-field approximation may significantly underestimate the achievable spectral efficiency of multi-user MIMO communications operating in the 30-GHz bands and above, even at ranges beyond the Fraunhofer distance.
For planar arrays with typical sizes, we show that computing combining schemes based on the far-field model significantly reduces the channel gain and spatial multiplexing capability. When the radiative near-field model is used, interference rejection schemes, such as the optimal minimum mean-square-error combiner, appear to be very promising, when combined with electrically large arrays, to meet the stringent requirements of next-generation networks.

\end{abstract}

%\vspace{-0.6cm}

\input{acronymsList}
\vspace{-0.5cm}

\section{Introduction and motivation}\label{sec:introduction}
%\IEEEPARstart{T}{o} file is intended to serve as a ``sample article file''
 \IEEEPARstart{T}{he} increasing demand for ubiquitous, reliable, fast, and scalable wireless services is pushing today's radio technology toward its ultimate limits. 
% The current deployment of fifth generation (5G) wireless networks is expected to exploit increasingly \ac{MIMO} techniques and cell densification in order to serve a large number of \ac{UE} per area with the required throughput. However, for sixth-generation (6G) wireless networks, even more stringent requirements are set. 
In this context, it is natural to continue searching for more bandwidth, which in turn pushes the operation towards higher frequencies~\cite{RappaportAccess2019}. 5G is designed to operate in bands up to $71$\,GHz \cite{rel17}. \Ac{THz} communications in the band from 0.1 to 10 THz is considered as a highly promising technology for 6G and beyond~\cite{RappaportAccess2019}. The use of high frequencies translates into higher path losses per antenna, which can be compensated for by antenna arrays. This combination undermines a fundamental assumption of multiple antenna communications: \emph{the wavefronts of radiated waves are locally planar over antenna arrays}~\cite{heath_lozano_2018}.

When an antenna radiates a wireless signal in free space, the wavefront of the electromagnetic waves has a different shape depending on the observation distance. Traditionally, two regions have been distinguished \cite{Selvan2017a}: the Fresnel and the far-field regions. Wireless communications have almost exclusively operated in the antenna (array) far field, which is conventionally characterized by propagation distances beyond the Fraunhofer distance. When arrays between $10$ cm and $1$ m are utilized, the typical communication ranges up to $100$ m are almost entirely in the Fresnel region when using a carrier frequency in the range $30$--$300\GHz$~\cite{LozanoMag2021}. Thus, the plane wave approximation does not hold anymore, and spherical wavefront propagation must be considered instead~\cite{LozanoMag2021}. This offers the opportunity for spatial-multiplexing in low-rank single-user \ac{MIMO} systems~\cite{bohagen09,Madhow2011} and for high-accuracy estimation of source position~\cite{Friedlander2019}. However, this line of research constitutes a minor fraction of the vast literature that relies on the plane-wave approximation.

\begin{figure}[t!]\vspace{-0.3cm}
  \begin{center}
    \begin{overpic}[width=\columnwidth]{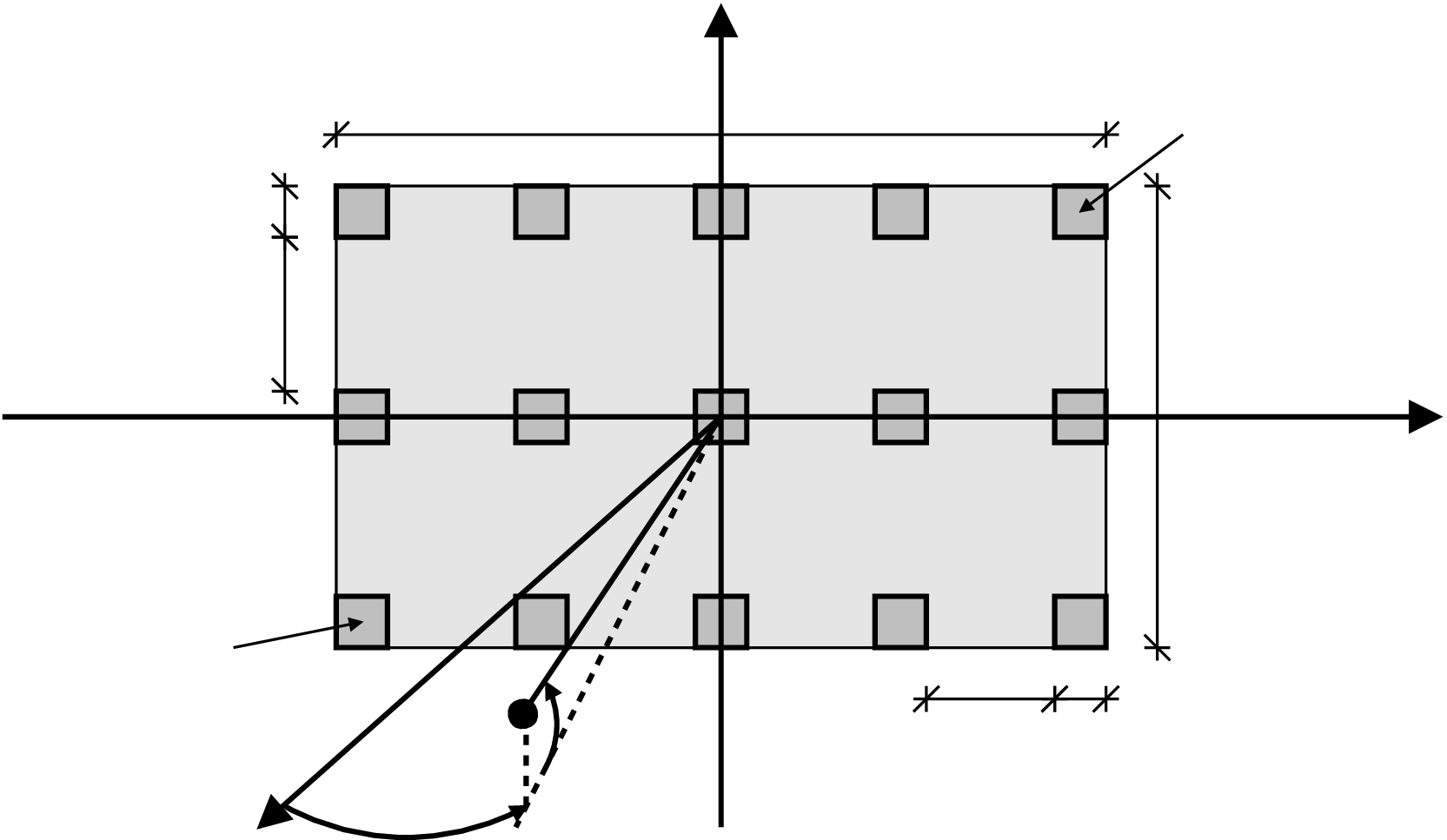}
      \put(45,50){\footnotesize{$L_\mathsf{H}$}}
      \put(81,25){{\footnotesize{$L_\mathsf{V}$}}}
      \put(66,06.5){\footnotesize{$d_\mathsf{H}$}}
      \put(71.5,06.0){\footnotesize{$\sqrt{A}$}}
      \put(15,33.5){\rotatebox{+90}{\footnotesize{$d_\mathsf{V}$}}}
      \put(15,41){\rotatebox{+90}{\footnotesize{$\sqrt{A}$}}}
      \put(31,07){\footnotesize{$\mathbf{s}_k$}}
      \put(25,02){\footnotesize{$\varphi_k$}}
      \put(39,05){\footnotesize{$\theta_k$}}
      \put(13,12){\footnotesize{$\mathbf{r}_1$}}
      \put(82,49){\footnotesize{$\mathbf{r}_N$}}
      \put(93,30){\footnotesize{$X$}}
      \put(45,54){\footnotesize{$Y$}}
      \put(19,05){\footnotesize{$Z$}}
    \end{overpic}
\caption{Diagram of the 2D planar array located in the $XY$-plane.}
\label{fig1}
\end{center}\vspace{-0.7cm}
\end{figure}

Our objective is to show that, in the bands above $6$~GHz, the classical far-field approximation may profoundly underestimate the achievable performance of multi-user \ac{MIMO} communication systems equipped with planar arrays of practical size, i.e., in the order of half-a-meter. The underestimation is already large in the mmWave band around $30\GHz$ bands, and is further exacerbated when higher frequencies are considered. Our numerical analysis also shows that, when the radiative near-field channel model is used, \ac{MMSE} combining vastly outperforms \ac{MR}, thanks to the sub-wavelength spatial resolution that largely increases its interference suppression capabilities. Particularly, \ac{MMSE} combining enables serving very many \acp{UE}; in the order of $1500$ \acp{UE}/km$^2$ per channel use (in line with 5G requirements \cite{imt2020}), while ensuring fairness across them (and thus significantly increasing the performance at the cell edge). This makes the combination of \ac{MMSE} combining and electrically large arrays a promising candidate to meet the stringent capacity requirements of next-generation networks.

\section{System and signal model}\label{sec:model}

We consider a planar array centered around the origin of the $XY$-plane, as shown in
\figurename~\ref{fig1}. The array consists of $N_\mathsf{V}$ horizontal rows and $N_\mathsf{H}$ antennas per row, for a total of $N=N_\mathsf{H}N_\mathsf{V}$
antennas. Each antenna has an area $A$
and the spacing is $d_\mathsf{H}$ and $d_\mathsf{V}$ along the horizontal and vertical directions, respectively. Thus, the horizontal and vertical lengths of the array are $L_\mathsf{H}= N_\mathsf{H}\sqrt{A}+\left(N_\mathsf{H}-1\right)d_\mathsf{H}$ and $L_\mathsf{V}= N_\mathsf{V}\sqrt{A}+\left(N_\mathsf{V}-1\right)d_\mathsf{V}$, respectively. 
The antennas are numbered from left to right and from the bottom row to the top row so that 
antenna $n$ is located at $\mathbf{r}_n=\left[x_n, y_n, 0\right]^\mathsf{T}$, where $x_n = \Delta_\mathsf{H}
  \left(-\frac{N_\mathsf{H}-1}{2}
     + \textrm{mod}\left(n-1, N_\mathsf{H}\right)
     \right)$ and $y_n = \Delta_\mathsf{V}
  \left(-\frac{N_\mathsf{V}-1}{2} + \left\lfloor{(n-1)}/{N_\mathsf{H}}\right\rfloor
  \right)$
% \begin{align}
%   \label{eq:x_antenna}
%   x_n &= \Delta_\mathsf{H}
%   \left(-\frac{N_\mathsf{H}-1}{2}
%      + \textrm{mod}\left(n-1, N_\mathsf{H}\right)
%      \right)\end{align}
%      \begin{align}
%   \label{eq:y_antenna}
%   y_n &= \Delta_\mathsf{V}
%   \left(-\frac{N_\mathsf{V}-1}{2} + \left\lfloor\frac{n-1}{N_\mathsf{H}}\right\rfloor
%   \right)
% \end{align}
with $\Delta_\mathsf{H}=\sqrt{A}+d_\mathsf{H}$ and
$\Delta_\mathsf{V}=\sqrt{A}+d_\mathsf{V}$.
We assume that $K$ single-antenna \acp{UE} communicate with the planar array depicted in \figurename~\ref{fig1} and transmit signals with
polarization in the $Y$ direction when traveling in the $Z$ direction~\cite{bjornson2020}.{\footnote{{The analysis can be extended to other polarization dimensions (e.g., linear combination of $X$ and $Y$ polarizations).}}} \Ac{LoS} propagation is considered, as it becomes predominant when considering high frequencies (and hence shrinking the transmission range) \cite{LozanoMag2021}. We denote by $\mathbf{s}_k=\left[x_k, y_k, z_k\right]^\mathsf{T}$ the arbitrary location
for source $k$ so that 
the signal impinges on the planar array with azimuth and elevation angles given by
$\varphi_k=\tan^{-1}({x_k}/{z_k})$ and
$\theta_k=\tan^{-1}({y_k}/{\sqrt{x_k^2+z_k^2}})$, respectively.

We let $\mathbf{h}_k=\left[h_{k1},\dots,h_{kN}\right]^\mathsf{T}\in\mathbb{C}^N$ denote the channel of
\ac{UE} $k$. In particular,
$h_{kn}=\left|h_{kn}\right|e^{-\mathsf{j}\phi_{kn}}$ is the channel from source $k$ to receive
antenna $n$, with $\left|h_{kn}\right|^2$ being the channel gain and $\phi_{kn}\in\left[0,2\pi\right)$
denoting the phase shift. In the remainder, perfect channel state information is assumed, since the channels $\left\{\mathbf{h}_k\right\}_{k=1}^{K}$ can be estimated arbitrarily well from pilot signals, thanks to the \ac{LoS} propagation.

\subsection{Channel model}\label{model:channel}
To model $\mathbf{h}_k$, we extend the prior work~\cite{bjornson2020}, which only considers the case $d_\mathsf{H} = d_\mathsf{V}= \sqrt{A}$. The extension to the case $d_\mathsf{H} \ne d_\mathsf{V}$ with $d_\mathsf{H},d_\mathsf{V} \ge \sqrt{A}$ is as follows.

  \begin{figure*}[t!]\vspace{-0.9cm}
  \setcounter{equation}{0}
  \begin{align}\label{eq:channelGain}
    \zeta_{kn} = 
    \frac{1}{12\pi} \sum_{i=0}^{1}{
      \sum_{j=0}^{1}{ \frac{ g_i\left( x_{kn} \right) g_j\left( y_{kn} \right) \left|z_k\right| }
        {\left( g_j^2\left( y_{kn} \right) + z_k^2 \right)
          \sqrt{ g_i^2\left( x_{kn} \right) + g_j^2\left( y_{kn} \right) + z_k^2 } } }
    }
    +& \frac{1}{6\pi} \sum_{i=0}^{1}{
      \sum_{j=0}^{1}{
        \tan^{-1}
        \left(
        \frac{ g_i\left( x_{kn} \right) g_j\left( y_{kn} \right) }
        { \left|z_k\right|
          \sqrt{ g_i^2\left( x_{kn} \right) + g_j^2\left( y_{kn} \right) + z_k^2 } }
        \right) } }
  \end{align}\vspace{-0.2cm}
  \hrule
    \end{figure*}
    
\begin{corollary}\label{lm:channelGain}
  Consider a lossless isotropic antenna located at $\mathbf{s}_k$ that transmits a signal that has
  polarization in the $Y$ direction when traveling in the $Z$ direction. The free-space channel gain
  $\zeta_{kn}$ at the $n$th receive antenna, located at $\mathbf{r}_n$ is given by \eqref{eq:channelGain}, shown at the top of the page,
  where 
  \setcounter{equation}{1}
  \begin{align}
  g_i\left( \alpha \right) \triangleq \sqrt{A}/2+\left(-1\right)^i \alpha
  \end{align}
  while $x_{kn} = x_k-x_n$ and $y_{kn} = y_k-y_n$.
  \hfill$ \blacksquare$
\end{corollary}

From \corollaryname~\ref{lm:channelGain}, the following model is obtained.
\begin{corollary}[Exact model]\label{cor:nearField}
  The channel entry $h_{kn}=\left|h_{kn}\right|e^{-\mathsf{j}\phi_{kn}}$ is obtained as
  \begin{align}\label{eq:nearField_modulus}
    \left|h_{kn}\right|&=\sqrt{\zeta_{k,n}}\\
    \label{eq:nearField_phase}
    \phi_{kn}&=2\pi\,\textrm{mod}\left(\frac{\left\|\mathbf{d}_{kn}\right\|}{\lambda}, 1\right)
  \end{align}
  where $\zeta_{kn}$ is defined in
  \eqref{eq:channelGain} and $\mathbf{d}_{kn}={\mathbf{s}_k}-{\mathbf{r}_n}$.
  \hfill$ \blacksquare$
\end{corollary}

The above model provides a general expression for $h_{kn}$ that allows to quantify its channel gain in the so-called
\emph{radiative near-field} of the array~\cite{dardari2020,bjornson2020}.\footnote{Throughout this
letter, we assume $\left\|\mathbf{d}_{kn}\right\|\gg\lambda$, so that
the system, although in the near-field region of the array, does not operate in the reactive near-field of the transmit antenna (see \cite{bjornson2020, dardari2020} for details).}
Since it captures the fundamental properties of wave propagation, we call it the \emph{exact model}. Notice that it is substantially different from the classical \emph{far-field model}, e.g.,~\cite{heath_lozano_2018}, that assumes locally planar wavefronts over arrays and is valid for distances beyond the Fraunhofer distance $d_F=2\left(L_\mathsf{H}^2+L_\mathsf{V}^2\right)/\lambda$ \cite[Eq.~(3)]{demir2021}.
%because it captures the three fundamental properties that makes it different from the far-field region, notably: \emph{i}) transmit-receive distance as a function of each array element; \emph{ii}) effective element area as a function of each per-element incident angle; and \emph{iii}) loss from polarization mismatch as a function of each per-element incident angle. For these reasons, we call it the \emph{exact model}. Notice that it is substantially different from the classical \emph{far-field model}, e.g.,~\cite{heath_lozano_2018}, that assumes locally plane wavefronts over antenna arrays and is valid for distances beyond the Fraunhofer distance $d_F$, obtained as $d_F=2\left(L_\mathsf{H}^2+L_\mathsf{V}^2\right)/\lambda$ \cite[Eq.~(3)]{demir2021}.

\begin{figure}[t]\vspace{-0.25cm}
\begin{center}
\includegraphics[width=\columnwidth]{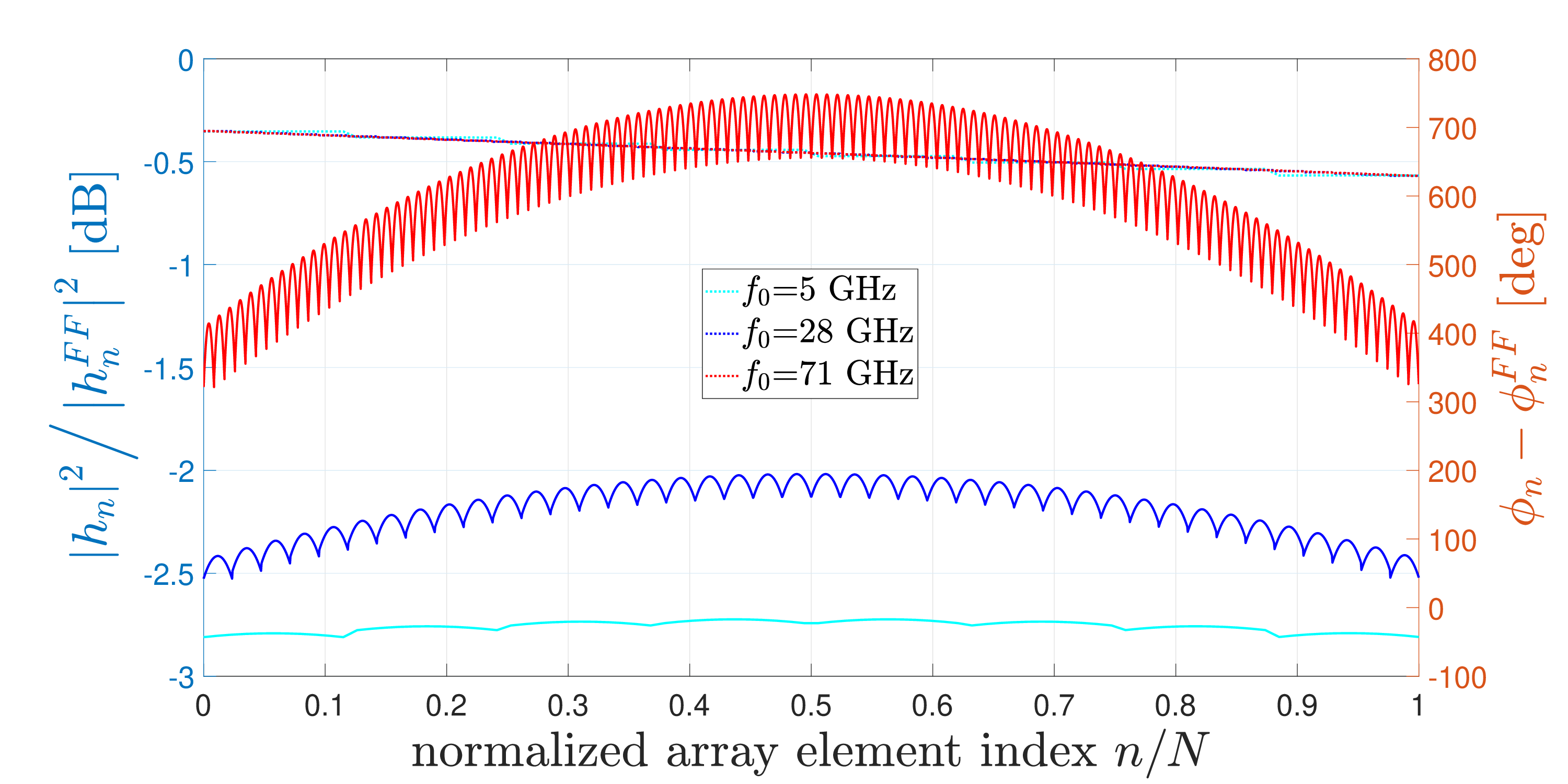}\vspace{-0.2cm}
\caption{Difference in the amplitude (left axis) and phase (right axis) between the exact model and the far-field approximation.}
\label{fig2}\vspace{-0.6cm}
\end{center}
\end{figure}

\begin{corollary}[Far-field approximation]\label{cor:farField}
  If \ac{UE} $k$ is in the far-field region of the array, i.e., 
  $d_k\cos\varphi_k\gg\max\left(L_\mathsf{H}, L_\mathsf{V}\right)$, then
  $h_{kn}\approx h_{kn}^{\rm FF}$ with $h_{kn}^{\rm FF} = \left|h_{kn}^{\rm FF}\right|e^{-\mathsf{j}\phi_{kn}}$ being modeled as
  \begin{align}\label{eq:farField_modulus}
    \left|h_{kn}^{\rm FF}\right|=\sqrt{
      \frac{A\cos\varphi_k}
      {4\pi d_k^2}}\\
    \label{eq:farField_phase}
    \phi_{kn}^{\rm FF}=\mathbf{k}^\mathsf{T}\left(\varphi_k, \theta_k\right)\mathbf{r}_n
  \end{align}
  where $\mathbf{k}\left(\varphi_k, \theta_k\right)=\frac{2\pi}{\lambda}
  \left[\cos\theta_k\sin\varphi_k, \sin\theta_k, \right.$
    $\left.\cos\theta_k\cos\varphi_k\right]^\mathsf{T}$
  is the wave vector, e.g.,~\cite{bjornson2017}.
  \hfill$ \blacksquare$
\end{corollary}
We notice that the propagation channel in \ac{MIMO} systems has been almost exclusively modeled as in \corollaryname~\ref{cor:farField}. For modern arrays of cellular networks\footnote{For instance, the Ericsson AIR 6419 product that contains $64$ antenna-integrated radios in a box that is roughly $1 \times 0.5\meter^2$ \cite{air6419}.}  of size $1 \times 0.5\meter^2$, this is a justified assumption when sub$-6$ GHz bands are used. In this case, $d_F \le 50\meter$ and, thus, most receivers are in the far-field of the transmitter.
%operating transmission ranges between $20\meter$ and $200\meter$ are almost entirely above the Fraunhofer region. 
The situation changes substantially in the frequency range $30$-$300\GHz$, in which $d_F \ge 250\meter$, and typical operating distances are entirely below it. This implies that the far-field approximation cannot be used, and the exact propagation model derived in \corollaryname~\ref{lm:channelGain} must be considered instead. As is known, the radiative near-field can create both noticeable amplitude variations and phase variations over the wavefront. To measure the impact of such variations, \figurename~\ref{fig2} reports the results for $L_\mathsf{H}=0.5\meter$, $L_\mathsf{V}=1.0\meter$, $A=\left(\lambda/4\right)^2$, $d_\mathsf{H}=0.5\lambda$, and $d_\mathsf{H}=2\lambda$, considering a \ac{UE} located at $30\meter$ from the \ac{BS}, which is elevated by $10\meter$. Amplitude variations are reported with dotted lines (using the left axis), whereas phase variations are represented by the solid lines (using the right axis). 
%Cyan and blue lines refers to the cases $f_0=5\GHz$ and $f_0=28\GHz$, respectively, whereas the red line refers to $f_0=71\GHz$. 
While the amplitude variations are negligible, the phase variations are  significant, particularly when the carrier frequency increases. 

Note that the model in \corollaryname~\ref{lm:channelGain} is also accurate in the far-field, thus there is no need to determine beforehand if the communication scenario is in the radiative near-field region or not. We conclude by noticing that the above discussion does not require the use of \emph{physically large arrays} (cf.~\cite{lu2021}), but holds true for commercially-sized arrays, e.g., in the order of half-a-meter wide and height. What matters is the size relative to the wavelength, the so-called \emph{electromagnetic size}.

\subsection{System model}\label{model:signal}

We consider the uplink.
The received signal is modeled as $
\mathbf{y}\in\mathbb{C}^N=\sum_{k=1}^{K}{\mathbf{h}_k s_k} + \mathbf{n}$,
where $s_k\sim\mathcal{N}_\mathbb{C}\left(0,p_k\right)$
is the data from \ac{UE} $k$ and
$\mathbf{n}\in\mathbb{C}^N$ is the thermal noise with i.i.d. elements distributed as $\mathcal{N}_\mathbb{C}\left(0,\sigma^2\right)$. To decode $s_k$, $\mathbf{y}$ is processed with the combining vector
$\mathbf{v}_k\in\mathbb{C}^N$. By treating the interference as noise, the \ac{SE} for \ac{UE} $k$ is
$\log_2\left(1+\gamma_k\right)$, where
\begin{align}\label{eq:sinr}
  \gamma_k = \frac{p_k \left|\mathbf{v}_k^\mathsf{H} \mathbf{h}_k\right|^2}
  { \sum_{i\neq k}{p_i \left|\mathbf{v}_k^\mathsf{H} \mathbf{h}_i\right|^2} + \sigma^2\left\|\mathbf{v}_k\right\|^2}
\end{align}
is the \ac{SINR}. We consider both \ac{MR} and \ac{MMSE} combining. {MR has low computational complexity and maximizes the power of the desired signal, but neglects interference. \ac{MMSE} has higher complexity but it maximizes the SINR in \eqref{eq:sinr}. Other suboptimal schemes, e.g. zero-forcing, are not considered for space limitation.} In the first
case, $\mathbf{v}_k^{\rm MR}=\mathbf{h}_k/\left\|\mathbf{h}_k\right\|$, while in the second case
\begin{align}\label{eq:mmse}
  \mathbf{v}_k^{\rm MMSE} = \left( \sum_{i=1}^{K}{{p_i \mathbf{h}_i \mathbf{h}_i^\mathsf{H} }} + {\sigma^2}\mathbf{I}_N\right)^{-1} \mathbf{h}_k
\end{align}
with $\mathbf{I}_N$ being the identity matrix of order $N$.

The vast majority of \ac{MIMO} literature for high frequencies (e.g., in the mm-Wave frequency bands) rely on the far-field approximation in \corollaryname~\ref{cor:farField} and, instead of estimating $\mathbf{h}_k$ directly, estimate the three parameters $\{d_k, \theta_k, \varphi_k\}$. The latter are used to obtain estimates of $\{\mathbf{h}^{\rm FF}_k; k=1,\ldots,K\}$ through~\eqref{eq:farField_modulus} and~\eqref{eq:farField_phase}, which are eventually used to compute the combiner. If the communication scenario is in the radiative near-field region, then the system operates inevitably in a \emph{mismatched mode}, no matter how good $\{d_k, \theta_k, \varphi_k\}$ have been estimated. From the above discussion, it thus follows that the combining vectors $\left\{\mathbf{v}_k\right\}_{k=1}^{K}$ can in practice follow either the \emph{exact model} defined in \corollaryname~\ref{cor:nearField} or the far-field approximation defined in \corollaryname~\ref{cor:farField}. The aim of this letter is to quantify the impact of such inaccurate channel modeling.

\section{The impact of a mismatched design}\label{sec:oneUser}

%To gain insights into the problem, we start considering basic scenarios with $K=1$ and $K=2$. 
We assume that the \ac{BS} is located at a height of $b=10\meter$. We further assume the following parameters, in line with the form factor of current 5G arrays: $L_\mathsf{H}=0.5\meter$, $L_\mathsf{V}=1.0\meter$, $A=\left(\lambda/4\right)^2$, $d_\mathsf{H}=0.5\lambda$ and $d_\mathsf{V}=2\lambda$. The communication takes places over a bandwidth of $B=100\MHz$, with the total receiver noise power $\sigma^2=-87\dBm$. Each \ac{UE} transmits with power $p_k=20\dBm$ $\forall k$. We assume a carrier frequency of $f_0=28\GHz$ such that $\lambda=10.71\mm$, $N_\mathsf{H}=62$, and $N_\mathsf{V}=42$, to focus on a 5G hot-spot scenario. When relevant, throughout the letter, we also consider higher carrier frequencies that cover future use cases and scenarios.

%and 2)~$f_0=71\GHz$ such that $\lambda=3.85\mm$, $N_\mathsf{H}=174$, and $N_\mathsf{V}=116$.

\subsection{Channel gain}
\begin{figure}[t]\vspace{-0.25cm}
  \begin{center}
    \begin{overpic}[width=1.1\columnwidth]{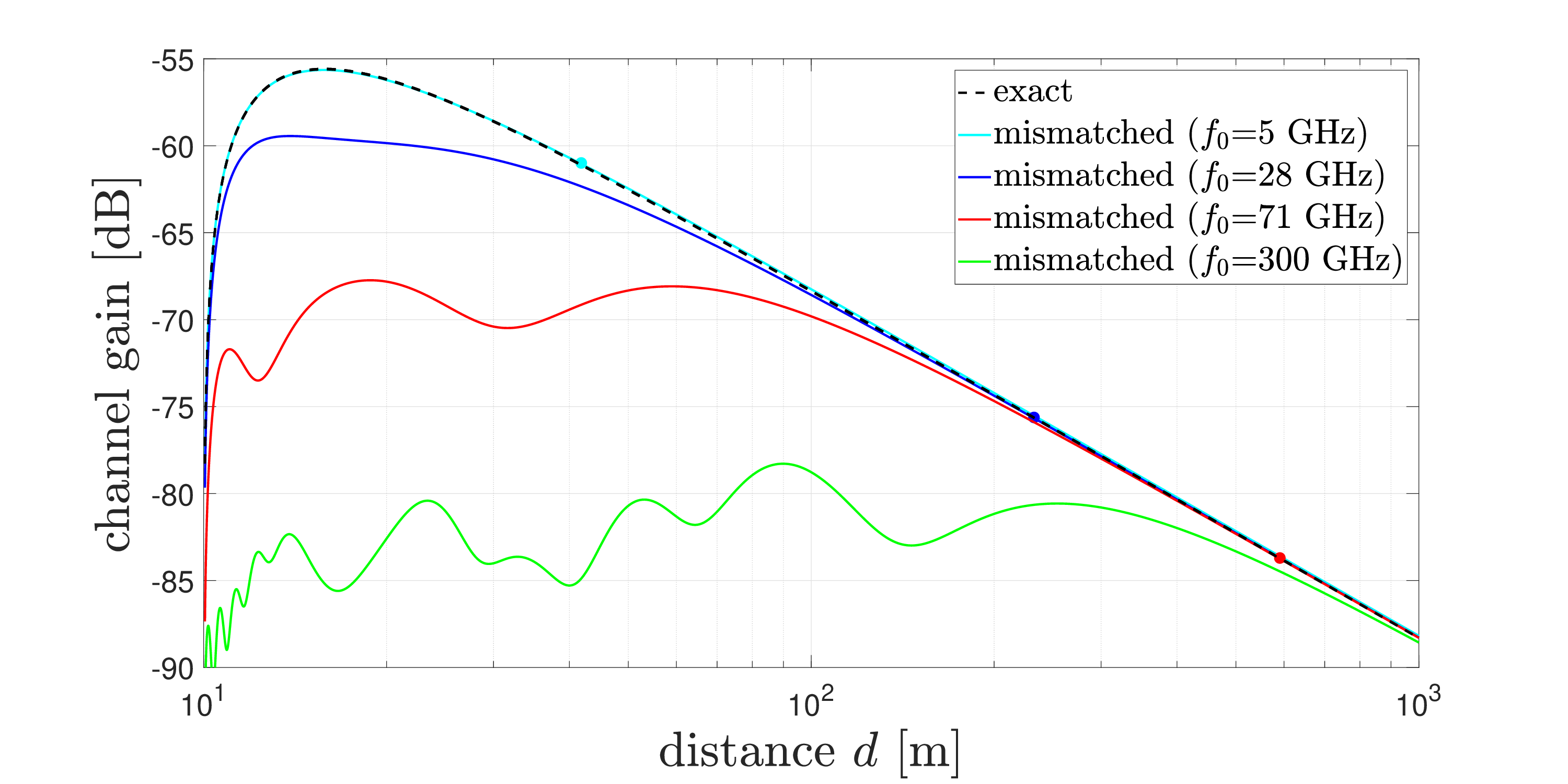}
      \put(74.5,27.5){\footnotesize{$d_F$}}
      \put(73.5,27){\vector(-2,-1){7}}
      \put(78.0,26){\vector(1,-3){3.5}}
      \put(72.5,28){\vector(-3,1){34}}
      %\put(50,23){\footnotesize{$d_B$}}
      %\put(50.8,25.6){\vector(-4,3){25}}
      %\put(52.5,25.5){\vector(1,3){1.3}}
      %\put(54.5,24){\vector(4,-1){16}}
    \end{overpic} 
\caption{Normalized channel gain $\left|\mathbf{v}_1^\mathsf{H} \mathbf{h}_1\right|^2/\left\|\mathbf{v}_1\right\|^2$ as a function of the distance of \acs{UE} $1$ along $Z$ axis.}
\label{fig3}
\end{center}\vspace{-0.6cm}
\end{figure}

We consider \ac{UE} 1 and assume that it is located along the $Z$ axis with coordinates
$\mathbf{s}_1=\left[0, -b, d\right]^\mathsf{T}$. In \figurename~\ref{fig3}, the dashed black line reports the normalized channel gain $\left|\mathbf{v}_1^\mathsf{H} \mathbf{h}_1\right|^2/\left\|\mathbf{v}_1\right\|^2$ achieved with the exact model, whereas the solid lines correspond to the channel gain measured with the combiners based on the far-field model. Particularly, the cyan and the blue lines refers to the cases $f_0=5\GHz$ and $f_0=28\GHz$, respectively, whereas the red and the green lines refer to $f_0=71\GHz$ and $f_0=300\GHz$, respectively. %Circular and square markers correspond to the Fraunhofer and Bj\"ornson distances, obtained as
Markers correspond to the Fraunhofer distances, obtained as
%$d_F=2\left(L_\mathsf{H}^2+L_\mathsf{V}^2\right)/\lambda$ and $d_B=2\left(L_\mathsf{H}^2+L_\mathsf{V}^2\right)\sqrt{N}$ \cite[Eqs.~(3) and (18)]{demir2021}. For $f_0=300\GHz$, $d_F=2.5\km$ falls outside the selected range.
$d_F=2\left(L_\mathsf{H}^2+L_\mathsf{V}^2\right)/\lambda$ \cite[Eq.~(3)]{demir2021}. For $f_0=300\GHz$, $d_F=2.5\km$ falls outside the selected range.

{We see that, thanks to the large values of $N$, the channel gain with the exact model depends very weakly on the carrier frequency (a difference of at most $0.3\dB$ between $5$ and $300$ GHz at the local maximum of the curve), and we thus only report one line, for clarity}. The same channel gain is achieved with the mismatched model only for sub-$6\GHz$ frequencies, irrespective of the distance. This validates the accuracy of the far-field approximation for such frequency bands. On the contrary, if higher frequencies are considered, then large differences are observed for transmission ranges below the Fraunhofer distance. This is a direct consequence of the inaccuracy of the far-field approximation in the Fresnel region. The gap increases as $f_0$ increases. Interestingly, we observe that, for transmission ranges of practical interest (up to a hundred of meters), it is significant already for $f_0 \ge 71\,$GHz. 
%In agreement with the findings from~\cite{demir2021}, the Bj\"ornson distance $d_B$ provides a tighter approximation. 
{Finally, note that the normalized channel gain is not monotonically decreasing as the distance from the \ac{BS} increases, but shows a local maximum. This is due to the specific choice of the \ac{BS} height $b$: when the \ac{UE} is too close to the \ac{BS}, the smaller path loss is overwhelmed by the loss due to the array directivity at larger elevations. Thus, the local maximum increases as $b$ increases.}

%This is not the cas, and shows a maximum at distance $d\approxeq16\meter$. This is a good tradeoff between spatial resolution of the array in the three-dimensional domain and the impact of path loss. We see that the channel gain is significantly different when the \ac{BS}-\ac{UE} distance is well below $d_F$ (the higher the carrier frequency, the higher the performance gap), while, as expected and in accordance with \cite{demir2021}, the two curves coincide for distances larger than $d_F$. Note that, since $A\propto\lambda^2$ and $N\propto 1/\lambda^2$ for a given array size, in the Franuhofer region $N\left|h_{kn}^{\rm FF}\right|^2$ does not depend on $\lambda$ (and hence the frequency $f_0$), and thus the blue and red curves coincide for a given distance $d$.\footnote{@LS: I don't know if it is worth mentioning it, the same thing should be said for the exact-model curve.}

\subsection{Interference gain}\label{sec:twoUsers}

%The scope of this section is to evaluate the impact of an interfering \ac{UE} in terms of \ac{SE} achieved in the uplink
%channel when using either a combining vector based on the exact model, or a combiner based on a mismatched model (i.e., the combining vector
%is based on the far-field approximation detailed in \corollaryname~\ref{cor:farField}). In this case, both \ac{MR} and \ac{MMSE} strategies are
%considered, and the channel gains follow the modeling detailed in \corollaryname~\ref{cor:nearField}.
We now analyze the normalized interference gain $\left|\mathbf{v}_1^\mathsf{H} \mathbf{h}_2\right|^2/\left\|\mathbf{v}_1\right\|^2$ with the exact and mismatched models. We assume \ac{UE} $1$ is placed at a fixed position along the $Z$ axis $\mathbf{s}_1=\left[0, -10\meter, +20\meter\right]^\mathsf{T}$, whereas the interfering \ac{UE} $2$ is transmitting from different locations $\mathbf{s}_2=\left[x_2, -10\meter, z_2\right]^\mathsf{T}$
over the $XZ$-plane (i.e., at the same height as \ac{UE} $1$). We assume $f_0=28\GHz$ and consider both \ac{MR} and \ac{MMSE}. Note that, at $f_0=28\GHz$, $d_F\approxeq233.3\meter$, and hence both \acp{UE} are in the near-field region.
%
%consider 
%The parameters for the planar array used by the \ac{BS} are the same used in \sectionname~\ref{sec:oneUser} considering $f_0=28\GHz$.
%The system is populated with $K=2$ \acp{UE}, modeled as follows. Both \acp{UE} transmit with power $p_k=20\dBm$, $k=1, 2$, whereas the
%total receiver noise power is equal to $-94\dBm$. \ac{UE} $1$ is placed at a fixed position along the $Z$ axis
%at a distance $10\meter$ ($\mathbf{s}_1=\left[0, -10\meter, +10\meter\right]^\mathsf{T}$), whereas \ac{UE} $2$ is transmitting from different locations
%over the $XZ$-plane (like \ac{UE} $1$, at $y_2=-b=-10\meter$).

\begin{figure}[t]\vspace{-0.7cm}
  \begin{center}
    \subfigure[\acs{MMSE} combining.]{
    \begin{overpic}[width=1.1\columnwidth]{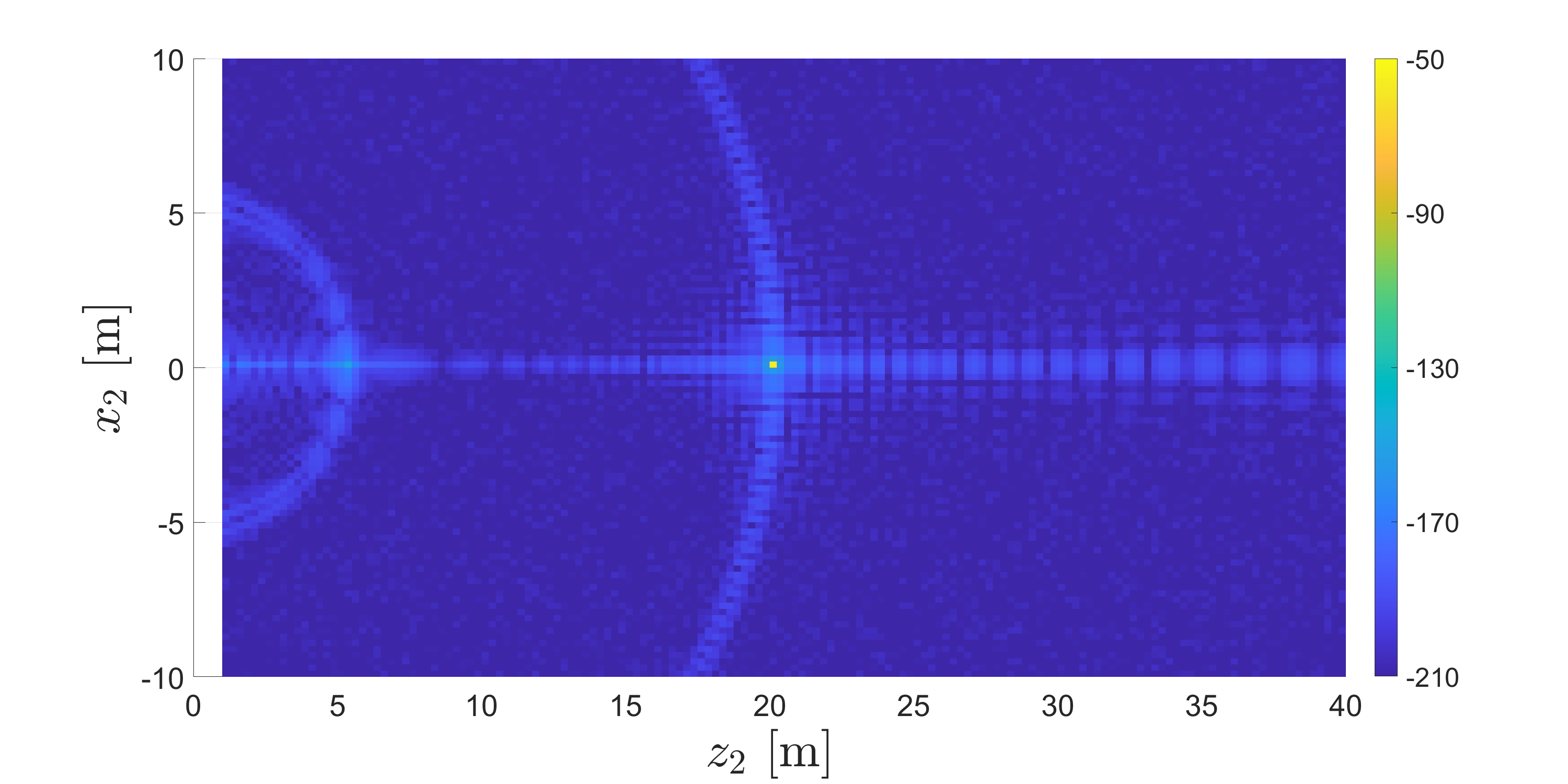}
      \put(56,34){\scriptsize{\textcolor{white}{\ac{UE} $1$'s position}}}
      \put(55,07){\frame{\includegraphics[scale=.16]{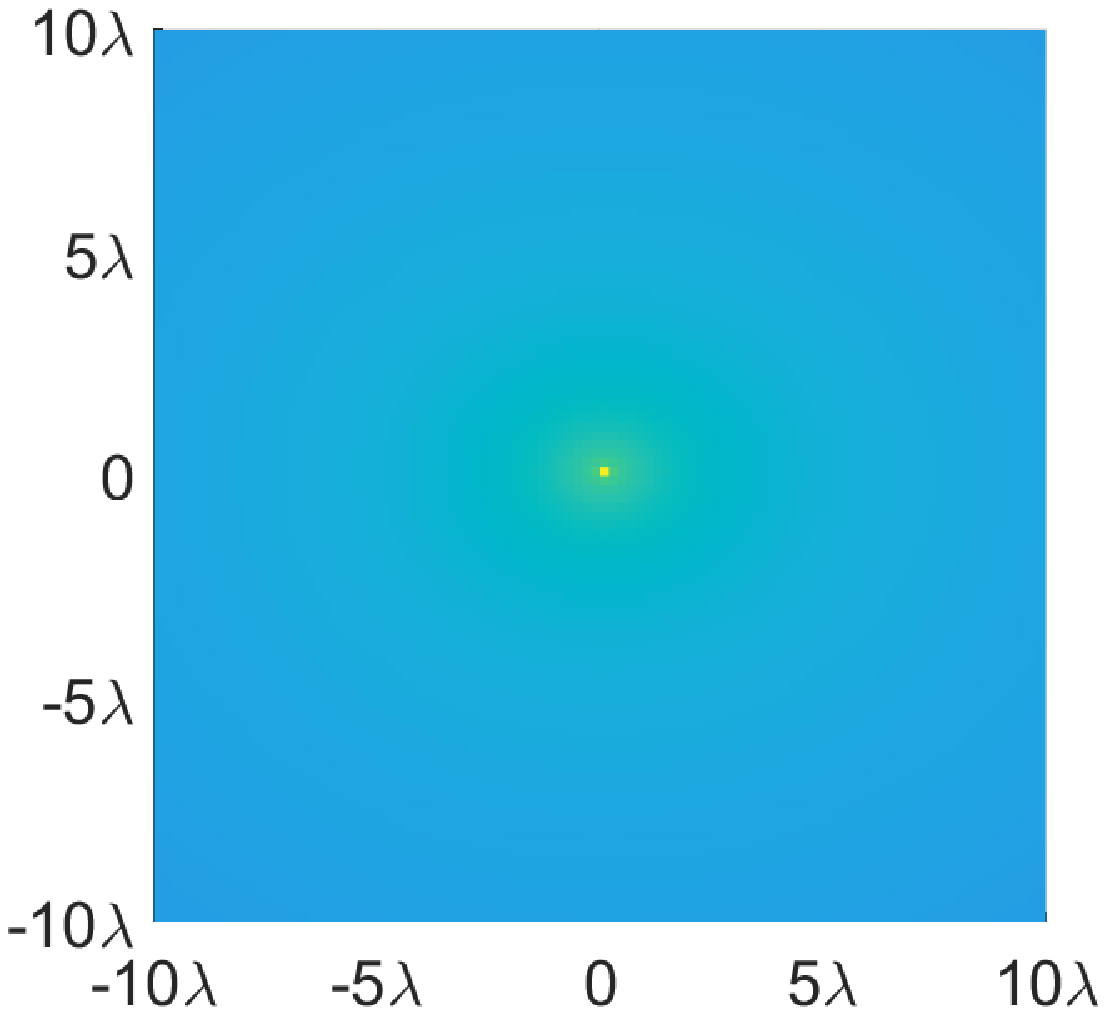}}}
      \put(59.5,33){\color{white}\vector(+1,-2){8}}
      \put(55.7,33){\color{white}\vector(-1,-1){6}}
    \end{overpic} 
    \label{fig4a}}\vspace{-0.2cm}
    \\
    \subfigure[\acs{MR} combining.]{
    \begin{overpic}[width=1.1\columnwidth]{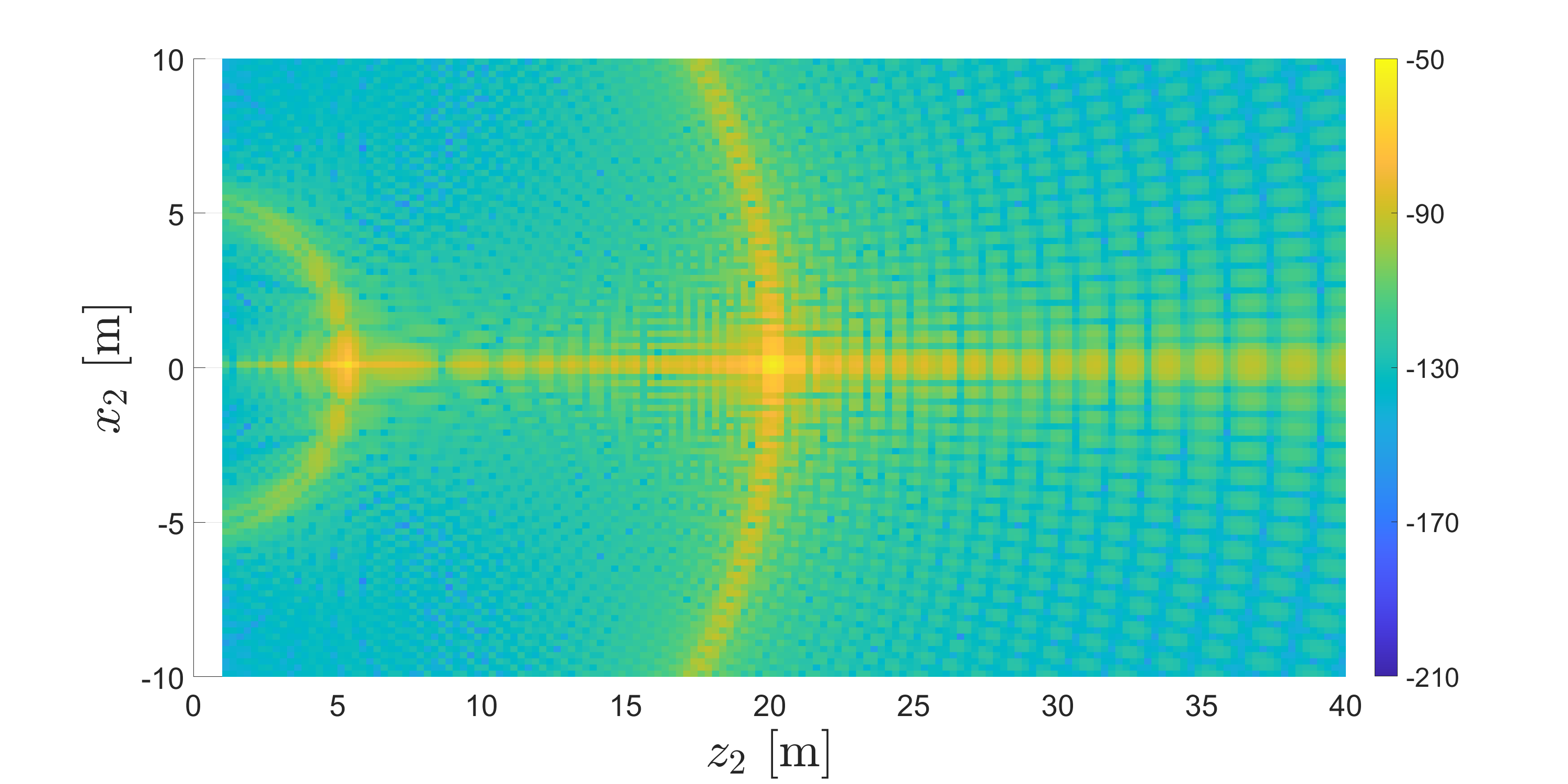}
      \put(56,34){\scriptsize{\textcolor{white}{\ac{UE} $1$'s position}}}
      \put(55,07){\frame{\includegraphics[scale=.16]{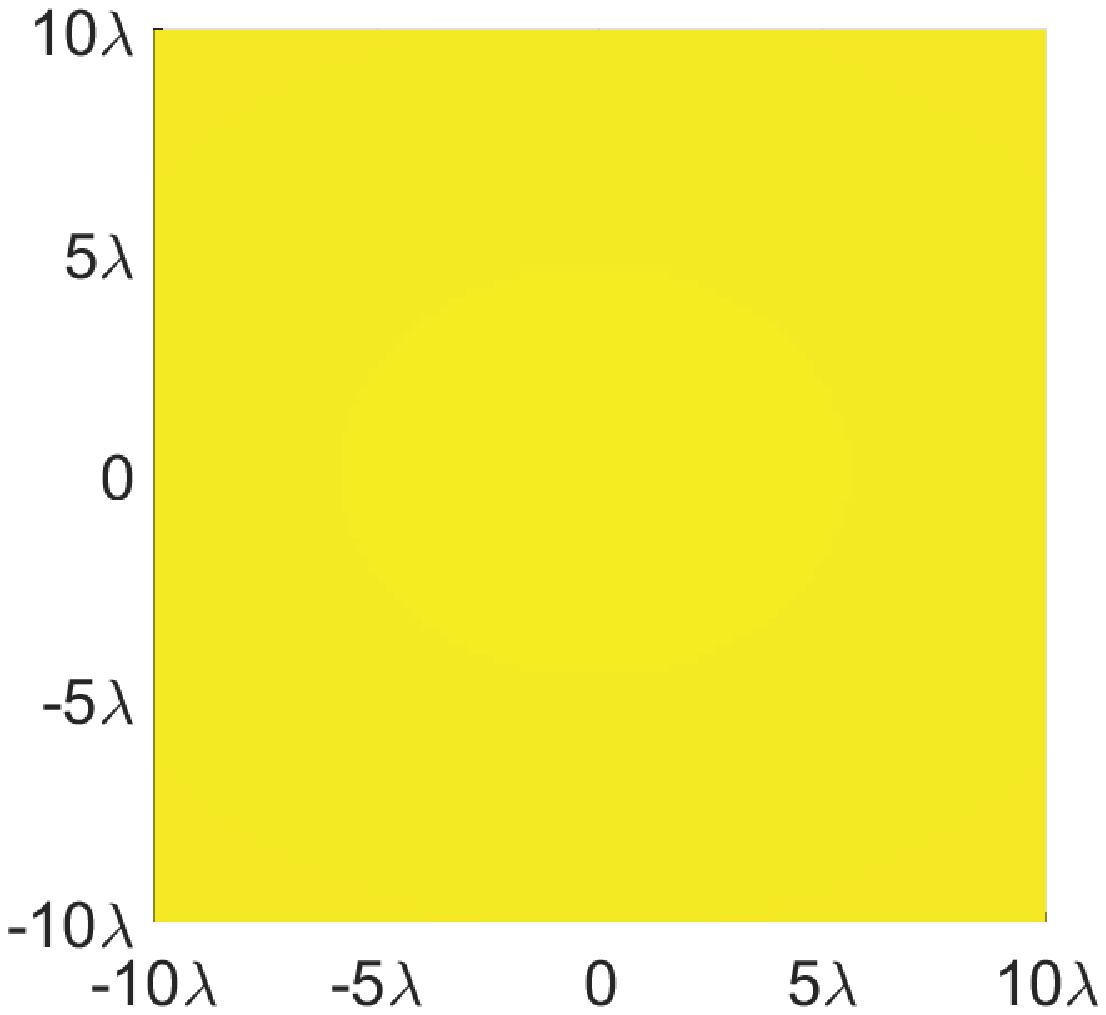}}}
      \put(59.5,33){\color{white}\vector(+1,-2){8}}
      \put(55.7,33){\color{white}\vector(-1,-1){6}}
    \end{overpic} 
    \label{fig4b}}
    \caption{Interference gain behavior using the \emph{exact} model for a fixed \acs{UE} as a function of different locations of an interfering \acs{UE}.}
    \label{fig4}
  \end{center}\vspace{-0.6cm}
\end{figure}

\figurename~\ref{fig4} reports the normalized interference gain with the exact model. \ac{MMSE} combining is used in \figurename~\ref{fig4}\subref{fig4a}, whereas the \ac{MR} is considered in \figurename~\ref{fig4}\subref{fig4b}. Each figure contains a magnification around \ac{UE} $1$'s location, in which the relative distance of \ac{UE} $2$ from \ac{UE} $1$ is measured in wavelengths (for a total span of around $1.07\times1.07\meter^2$).  \figurename~\ref{fig4}\subref{fig4a} shows that the interference with \ac{MMSE} is high only in a small region around \ac{UE} $1$, whose semi-major axis (along both directions) is fractions of the wavelength. This means that \ac{MMSE} can efficiently reject any interfering signal that comes from a location that is at least a few centimeters away. On the contrary, \figurename~\ref{fig4}\subref{fig4b} shows that \ac{MR} experiences high interference from locations that are either along the $Z$ direction or along a semi-circle with radius equal to the distance of \ac{UE} $1$.

%On the contrary, the results of \figurename~\ref{fig4}\subref{fig4b} are remarkably different. As can be seen, the lack of \ac{MAI} suppression of the \ac{MR} criterion leads to much poorer performance compared to the \ac{MMSE} counterpart in terms of spatial resolution, and the results provided by the combiner based on the exact model are comparable with those using the far-field approach. In particular, the curve shows many fluctuations (with maxima along a semi-circle with radius equal to the fixed-user distance from the \ac{BS} and along the line connecting the \ac{BS} with \ac{UE} $1$'s location), while the near-field-based curve is nearly flat, thus representing a huge difference between the two approaches.

%\begin{figure}[t]
%  \begin{center}
%    \subfigure[exact model.]{
%    \begin{overpic}[width=\columnwidth]{figures/twoUsers_interferenceGain_MR_fixedUser_exact_low}
%      \put(56,33){\scriptsize{\textcolor{white}{\ac{UE} $1$'s position}}}
%      \put(55,32){\color{white}\vector(-1,-1){6}}    \end{overpic} 
%    \label{fig4b}}
%    \\
%    \subfigure[mismatched model.]{
%    \begin{overpic}[width=\columnwidth]{figures/twoUsers_interferenceGain_MR_fixedUser_mismatched_low}
%      \put(56,33){\scriptsize{\textcolor{white}{\ac{UE} $1$'s position}}}
%      \put(55,32){\color{white}\vector(-1,-1){6}}    \end{overpic} 
%    \label{fig5b}}
%    \caption{Interference gain behavior with \acs{MR} for a fixed user as a function of different locations of an interfering user.}
%    \label{fig:twoUsers_MR_fixedUser}
%  \end{center}
%\end{figure}

\begin{figure}[t]\vspace{-0.7cm}
  \begin{center}
    \subfigure[\acs{MMSE} combining.]{
    \begin{overpic}[width=1.1\columnwidth]{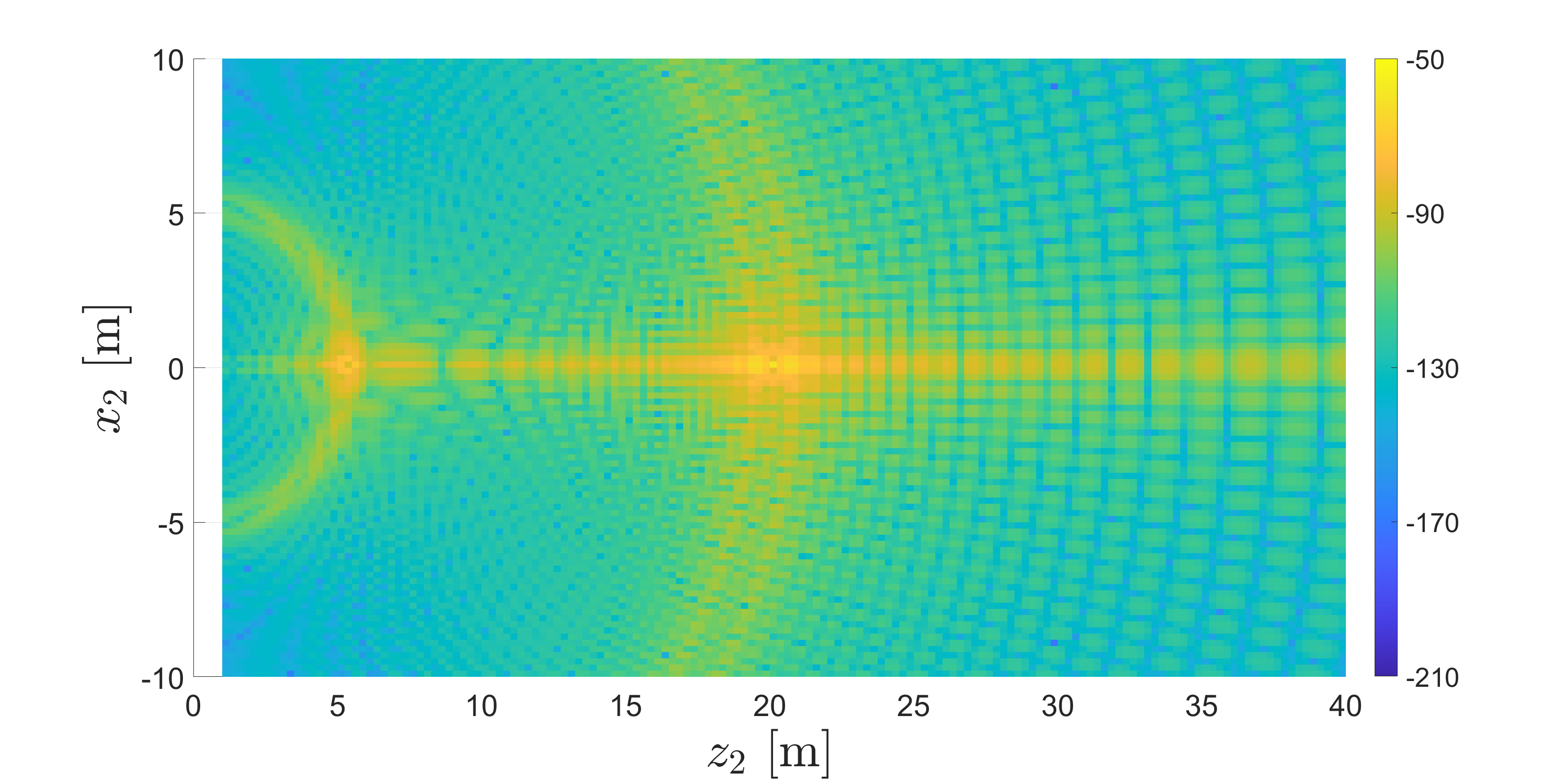}
      \put(56,34){\scriptsize{\textcolor{white}{\ac{UE} $1$'s position}}}
      \put(55,07){\frame{\includegraphics[scale=.16]{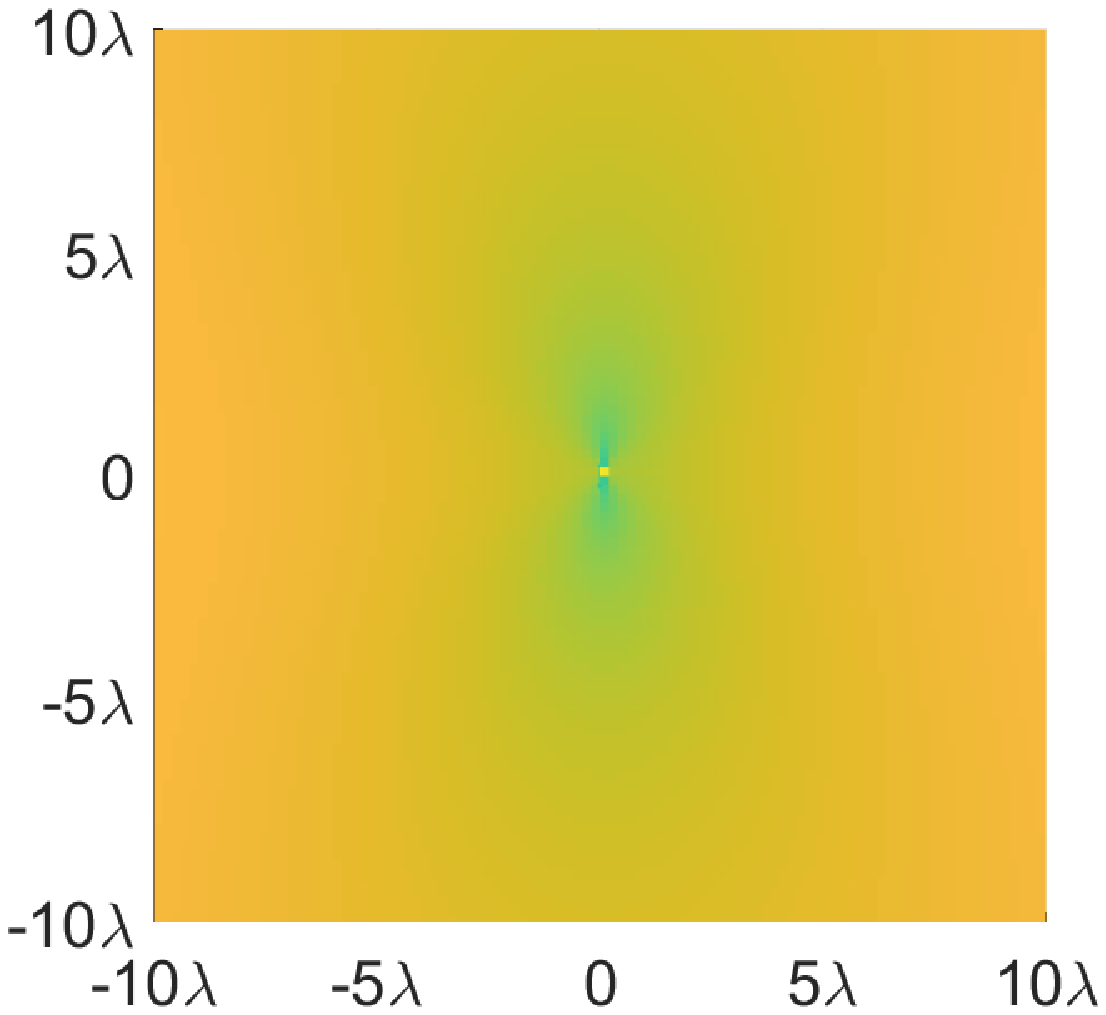}}}
      \put(59.5,33){\color{white}\vector(+1,-2){8}}
      \put(55.7,33){\color{white}\vector(-1,-1){6}}
    \end{overpic} 
    \label{fig5a}}\vspace{-0.2cm}
    \\
    \subfigure[\acs{MR} combining.]{
    \begin{overpic}[width=1.1\columnwidth]{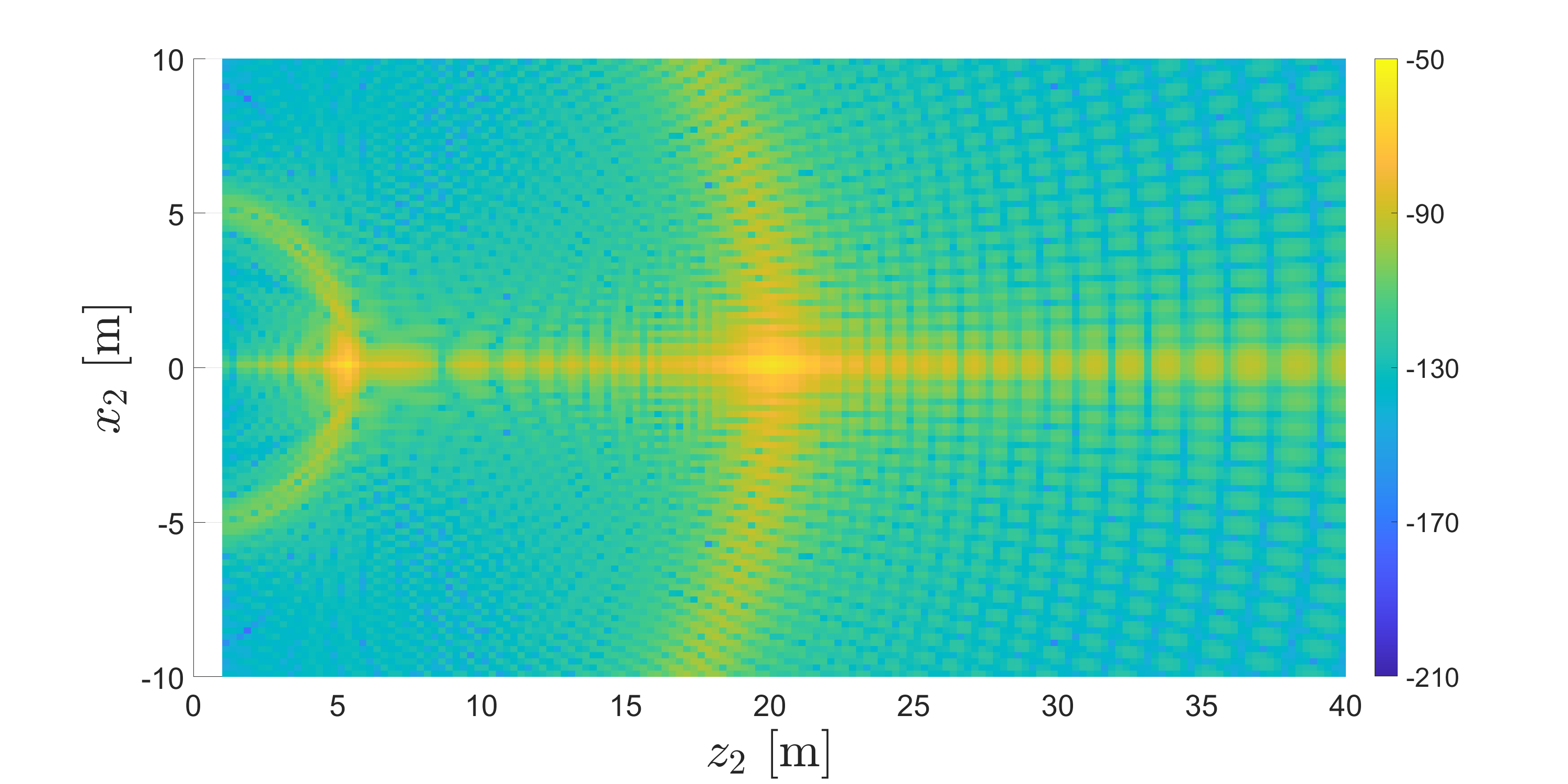}
      \put(56,34){\scriptsize{\textcolor{white}{\ac{UE} $1$'s position}}}
      \put(55,07){\frame{\includegraphics[scale=.16]{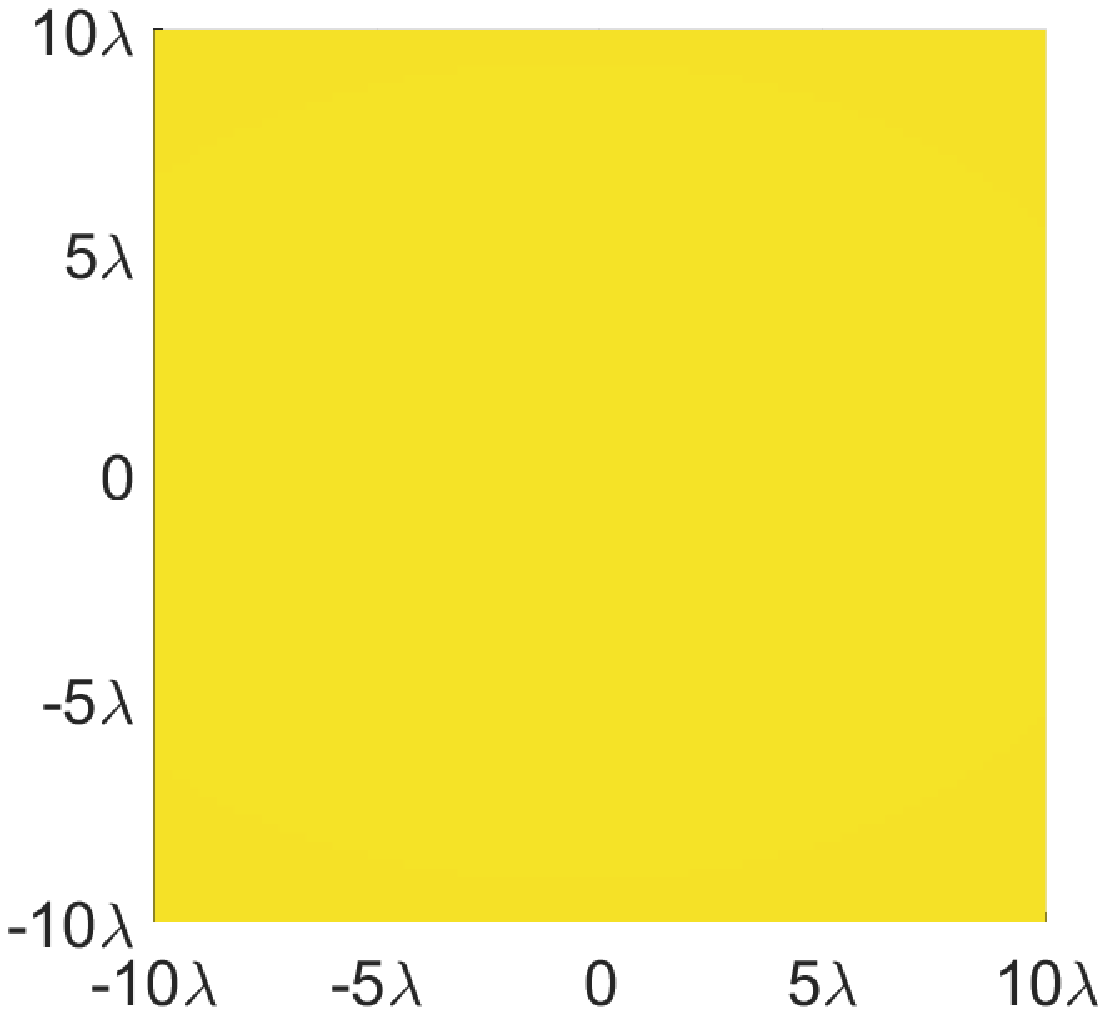}}}
      \put(59.5,33){\color{white}\vector(+1,-2){8}}
      \put(55.7,33){\color{white}\vector(-1,-1){6}}
    \end{overpic} 
    \label{fig5b}}
    \caption{Interference gain behavior using the \emph{mismatched} model for a fixed \acs{UE} as a function of different locations of an interfering \acs{UE}.}
    \label{fig5}
  \end{center}\vspace{-0.6cm}
\end{figure}

%Analogously to \figurename~\ref{fig:twoUsers_MMSE_fixedUser}\subref{fig4a}, \figurename~\ref{fig:twoUsers_MMSE_movingUser}\subref{fig:twoUsers_MMSE_movingUser_exact} shows that user $2$'s interference gain is increased by the interference produced by the fixed user $1$ only when their relative distance is within fractions of the wavelength $\lambda$. Interestingly, values of user $2$'s \ac{SE} are larger when it is placed close to the \ac{BS} (at $y_2=-b$), due to the poor resolution offered by the planar array when the \ac{UE} is below it, and follow the path loss experienced when \ac{UE} $2$ moves at various locations in the $XZ$ plane. On the other hand, when inspecting the results of \figurename~\ref{fig:twoUsers_MMSE_movingUser}\subref{fig:twoUsers_MMSE_movingUser_mismatched}, the behavior is almost similar to the one observed in \ref{fig:twoUsers_MMSE_fixedUser}\subref{fig5b} for the fixed user's case, and most importantly, the difference in the values with respect to \figurename~\ref{fig:twoUsers_MMSE_fixedUser}\subref{fig4a} is very large in almost all locations.

\figurename~\ref{fig5} plots the results obtained with the mismatched model. Unlike \figurename~\ref{fig4}, we now see that the interference gain with the two combining strategies exhibits a similar behaviour. The impact of the mismatched model is particularly evident with the \ac{MMSE} combiner. In particular, we observe that the lower values of the interference gain are at least three orders of magnitude higher than those in \figurename~\ref{fig4} (note that the same colorbar scale is used in all figures, including the magnifications). The conclusion is that \ac{MMSE} can suppress interference much more efficiently  when used with the exact model, whereas \ac{MR} is greatly suboptimal in both cases.

%\subref{fig4a}, and available only when the interfering user is placed meters away from the fixed one. This means that using an \ac{MMSE} combining strategy coupled with the exact, matched model yields a significant performance gain in terms of interference suppression provided by either a mismatched-based or an \ac{MR}-based combining strategy.

%% \begin{figure}[h!]
%%   \begin{center}
%%     \subfigure[exact model.]{
%%     \begin{overpic}[width=\columnwidth]{figures/twoUsers_MR_movingUser_exact}
%%      \put(55,28){\scriptsize{\ac{UE} $1$'s position}}
%%      \put(54,27){\vector(-1,-1){6}}
%%     \end{overpic} 
%%     \label{fig:twoUsers_MR_movingUser_exact}}
%%     \\
%%     \subfigure[mismatched model.]{
%%     \begin{overpic}[width=\columnwidth]{figures/twoUsers_MR_movingUser_mismatched}
%%      \put(55,28){\scriptsize{\ac{UE} $1$'s position}}
%%      \put(54,27){\vector(-1,-1){6}}
%%     \end{overpic} 
%%     \label{fig:twoUsers_MR_movingUser_mismatched}}
%%     \caption{\acs{SE} behavior with \acs{MR} for a moving user as a function of its locations (considering a fixed interfering user).}
%%     \label{fig:twoUsers_MR_movingUser}
%%   \end{center}
%% \end{figure}

\subsection{Spectral efficiency analysis}\label{sec:multipleUsers}

%To gain insights into the problem, we start considering basic scenarios with $K=1$ and $K=2$. We assume that the \ac{BS} is located at a height of $b=10\meter$. Also, we consider $L_\mathsf{H}=0.5\meter$, $L_\mathsf{V}=1.0\meter$, $A=\left(\lambda/4\right)^2$, $d_\mathsf{H}=0.5\lambda$ and $d_\mathsf{H}=2\lambda$. The communication takes places over a bandwidth of $B=20\MHz$,
%with a total receiver noise power $\sigma^2=-94\dBm$. Each \ac{UE} transmits with power $p_k=20\dBm$ $\forall k$. We assume: 1) $f_0=28\GHz$ such that $\lambda=10.71\mm$, $N_\mathsf{H}=62$, and $N_\mathsf{V}=42$; and 2)
%$f_0=71\GHz$ such that $\lambda=3.85\mm$, $N_\mathsf{H}=174$, and $N_\mathsf{V}=116$.
We now evaluate the \ac{SE} of a single-cell network, assuming that $K$ \acp{UE} are randomly displaced in the sector $\left[-\pi/3, +\pi/3\right)$ with a minimum distance of $15\meter$ from \ac{BS}. 
%The numerical results are obtained from $100$ independent simulations.

\begin{figure}[t]\vspace{-0.7cm}
  \begin{center}
    \subfigure[\acs{MMSE} combining.]{
    {\includegraphics[width=\columnwidth]{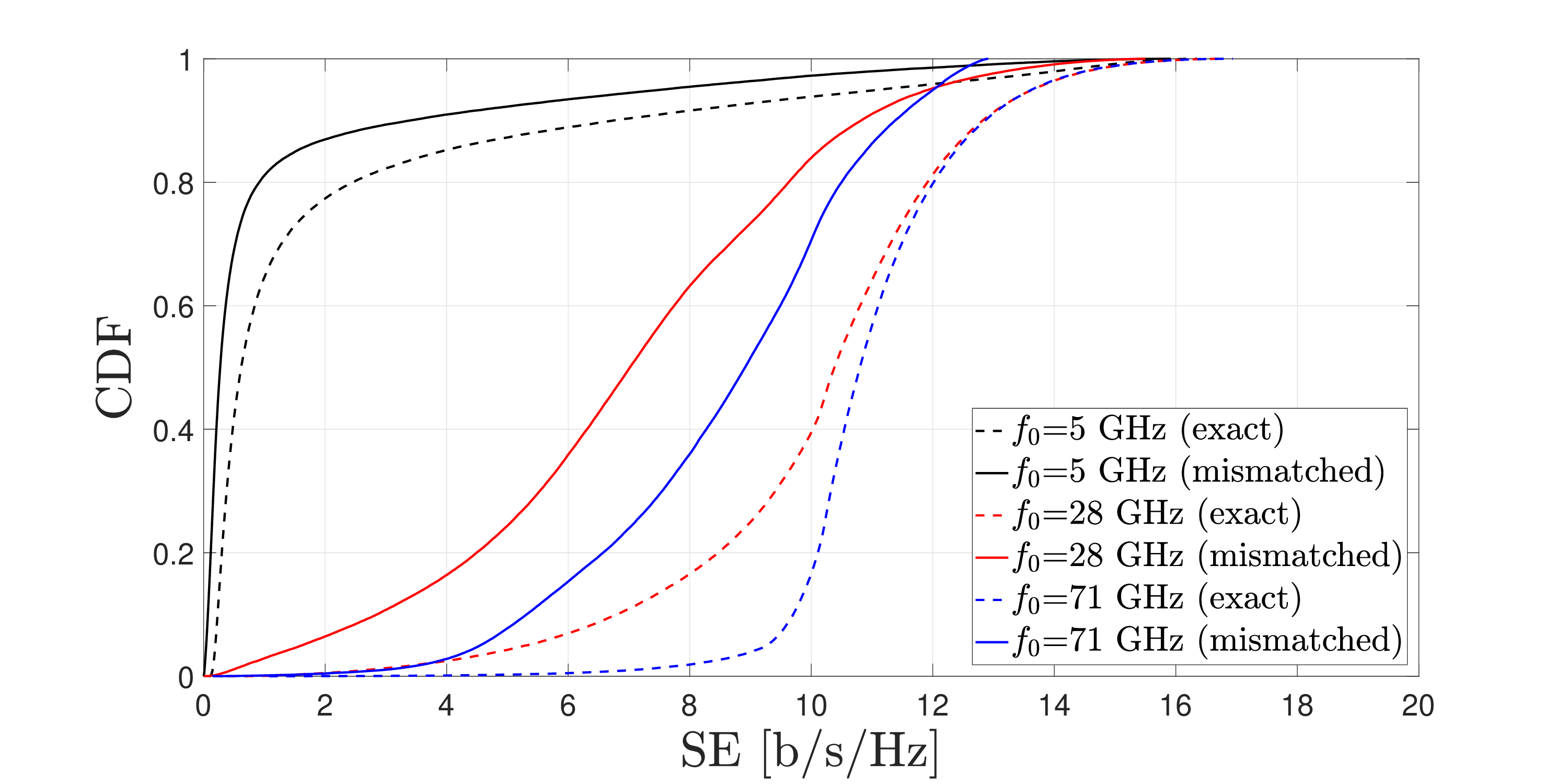}}
    \label{fig6a}}\vspace{-0.3cm}
    \\
    \subfigure[\acs{MR} combining.]{
    {\includegraphics[width=\columnwidth]{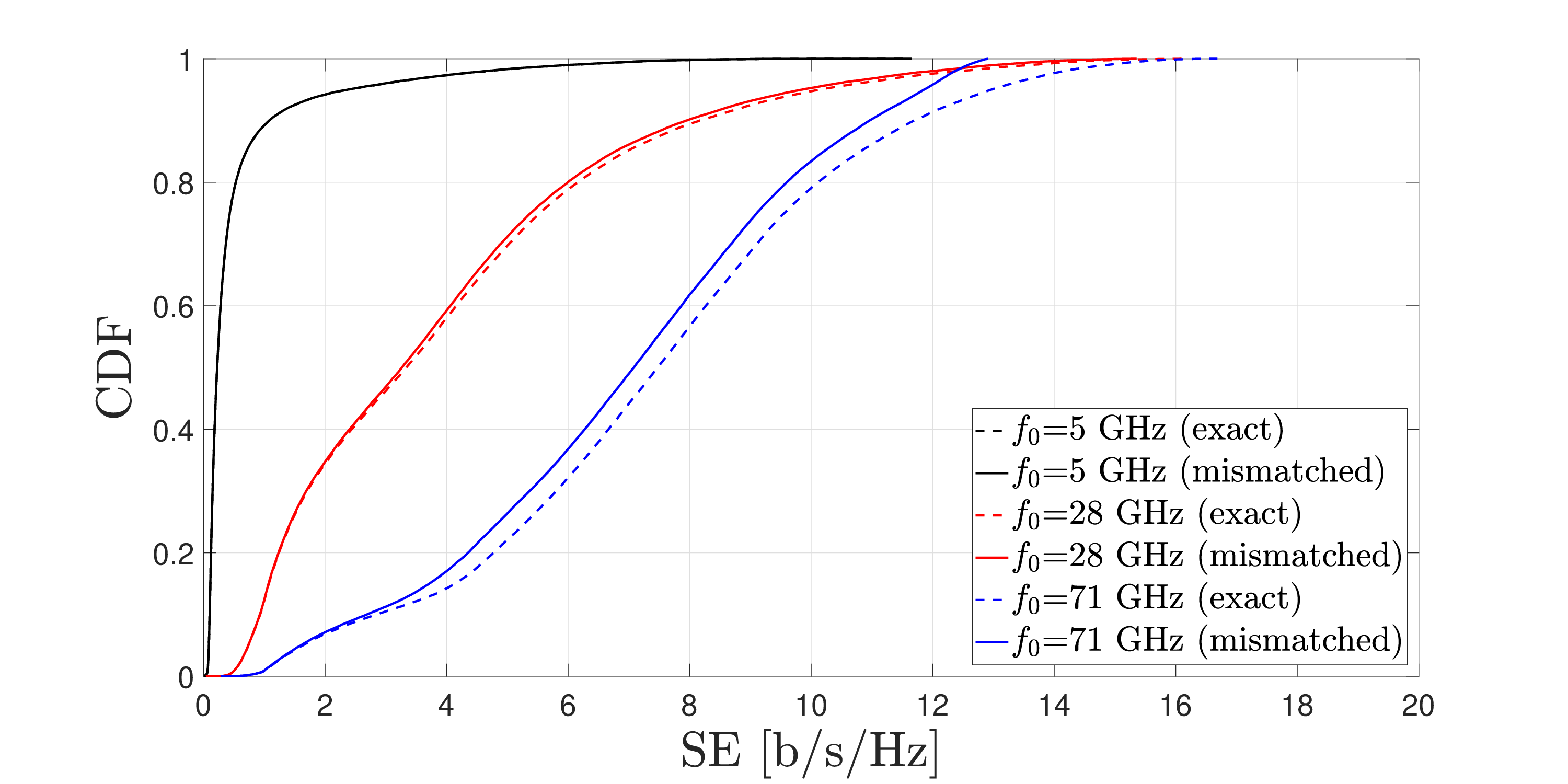}} 
    \label{fig6b}}
    \caption{\acs{CDF} of the \acs{SE} as a function of the carrier frequency when $K=100$ and $R = 230$\,m, corresponding to the Fraunhofer distance at $28\GHz$.}
    \label{fig6}
  \end{center}\vspace{-0.5cm}
\end{figure}

\figurename~\ref{fig6} reports the \ac{CDF} of the \ac{SE} achieved by \ac{MMSE} and \ac{MR} when the carrier frequency is $f_0=5\GHz$, $28\GHz$, and $71\GHz$. The number of \acp{UE} is $K=100$, and the cell radius is $R=230\meter$, which corresponds approximately to the Fraunhofer distance at $28\GHz$, and is in line with current 5G cell sizes. The dashed and solid refer to the results obtained with the exact and mismatched model, respectively. Let us focus on the results of \figurename~\ref{fig6}\subref{fig6a}, obtained with the \ac{MMSE} combiner. For all frequencies, the performance using the exact model to build the combiner is much better than the one obtained with the mismatched model. However, increasing the carrier frequency has a two-fold beneficial impact on the \ac{SE}. On the one hand, the average \ac{SE} increases as $f_0$ increases. On the other hand, the \ac{CDF} with the exact model exhibits a steeper behavior, meaning that more fairness across the \ac{UE} positions is guaranteed. The conclusion is that the use of the exact model dramatically improves the interference suppression capabilities of \ac{MMSE} combining. This is not the case with \ac{MR} combining. Indeed, \figurename~\ref{fig6}\subref{fig6b} shows that only marginal differences exist between exact and mismatched models, including the fairness properties.

\begin{figure}[t]\vspace{-0.7cm}
  \begin{center}
    \subfigure[\acs{CDF} of the \acs{SE} with \acs{MMSE} combining for different values of $K$.]{
    {\includegraphics[width=\columnwidth]{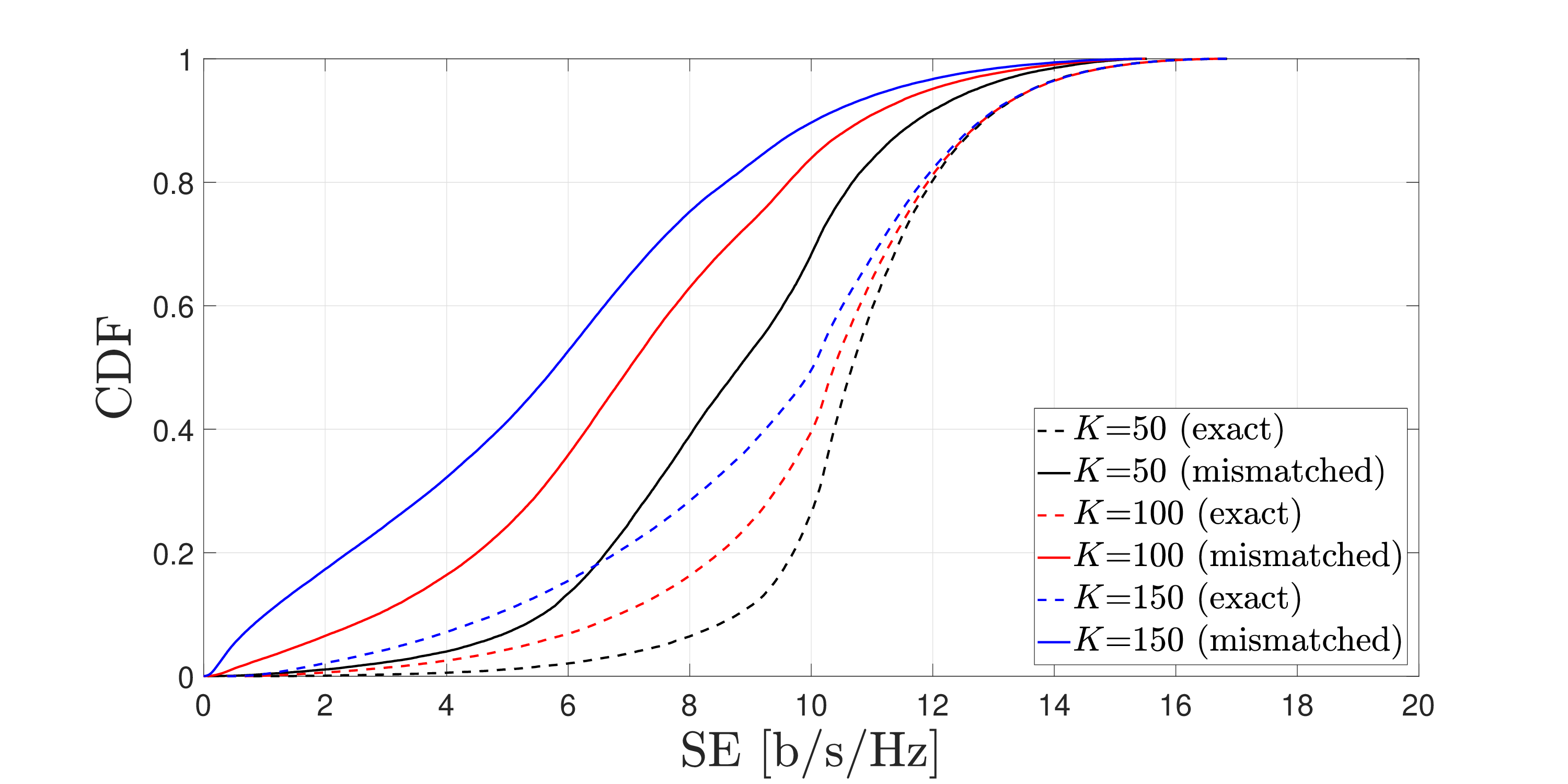}}
    \label{fig7a}}\vspace{-0.3cm}
    \\
    \subfigure[\acs{SE} per \acs{UE} of both \acs{MMSE} and \acs{MR} as a function of $K$.]{
    {\includegraphics[width=\columnwidth]{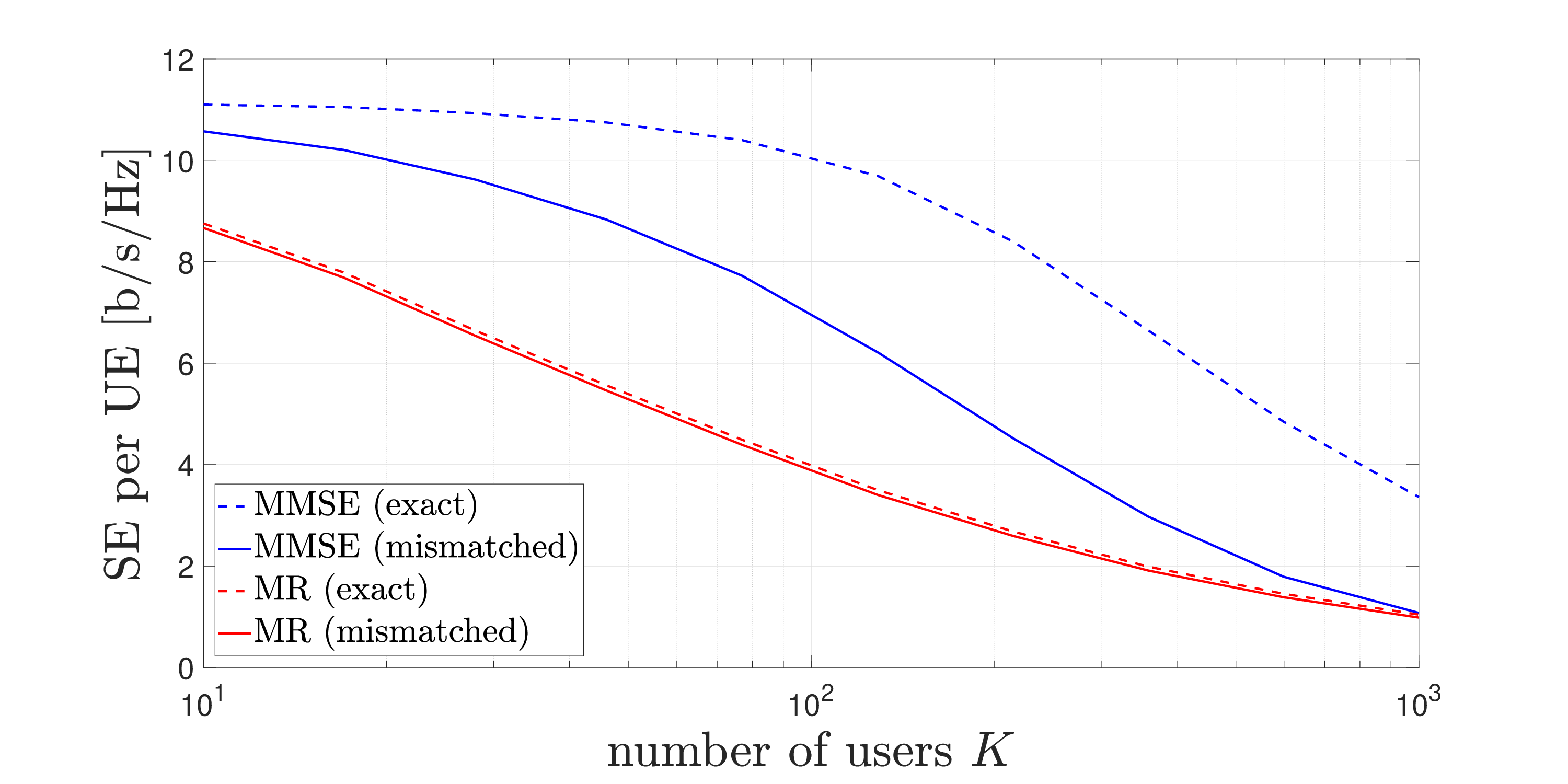}} 
    \label{fig7b}}
    \caption{Impact of the number of active \acp{UE} on the \acs{SE} performance.}
    \label{fig7}
  \end{center}\vspace{-0.6cm}
\end{figure}

A similar conclusion can be drawn in \figurename~\ref{fig7}\subref{fig7a}, which collects the \acp{CDF} for the \ac{SE} achieved by \ac{MMSE} and \ac{MR} combiners for a variable number of \acp{UE} ($K=50$, $100$, and $150$) when $f_0=28\GHz$ and $R=230\meter$. We notice that the results are only marginally affected by the number of \acp{UE} when the exact model is used to build the combiner (dashed lines). This is not true for the combiners based on the mismatched model (solid lines) and/or based on  \ac{MR} combining (not reported here for the sake of brevity, as, similarly to \figurename~\ref{fig6}\subref{fig6b}, both models provide very similar results when using \acs{MR} combining). The same conclusion applies on the fairness performance, which is guaranteed by the usage of \ac{MMSE} combining based on the exact model.

%The system parameters are as follows: the \ac{BS} is equipped with the planar rectangular antenna array modeled
%in \sectionname~\ref{sec:model}, using the same parameters listed in \sectionname~\ref{sec:oneUser} for $f_0=28\GHz$.
%All $K$ users in the system transmit with power $p_k=20\dBm$, $k=1, \dots, K$, whereas the
%total receiver noise power is equal to $-94\dBm$.

%\begin{figure}[t]
%\begin{center}
%{\includegraphics[width=\columnwidth]{figures/fig8}}
%\caption{Average per-\acs{UE} sum \acs{SE} as a function of the cell radius $R$.}
%\label{fig8}
%\end{center}
%\end{figure}

To further quantify the benefits brought by \ac{MMSE} combining with the exact model, \figurename~\ref{fig7}\subref{fig7b}  provides the average \ac{SE} per \ac{UE} as a function of $K$. The cell sector radius $R$ is $230\meter$, and the carrier frequency is $f_0=28\GHz$. As can be seen, the gap between the exact and mismatched models is very significant. Moreover, the \ac{SE} maintains nearly flat as $K$ increases up to $K=100$, which translates into a density of around $1500$ devices/km$^2$ per channel use, as requested in 5G \cite{imt2020}. Note that the excellent interference suppression capabilities let \ac{MMSE} based on the exact model maintain the gap with other schemes when $K$ reaches very large values. The same trend is confirmed when measuring the average \ac{SE} per \ac{UE} as a function of $R$ (not reported for the sake of brevity): the performance gap using current 5G frequencies is very significant for $R\le50\meter$, and even more significant when considering sub-THz frequencies and smaller cell sizes.

%\begin{figure}[t]
%\begin{center}
%{\includegraphics[width=\columnwidth]{figures/multipleUsers_vs_arrayHeight}}
%\caption{Average per-\acs{UE} sum \acs{SE} behavior as a function of \acsp{BS} height $b$.}
%\label{fig:multipleUsers_vs_arrayHeight}
%\end{center}
%\end{figure}

\section{Conclusion}\label{sec:conclusion}

{The main conclusion of this letter is that it is time for multi-user \ac{MIMO} communication theorists to abandon the far-field approximation when considering carrier frequencies above $6$~GHz. We instead need to consider more complicated channel models that capture the radiative near-field characteristics, in particular concerning the spherical phase variations. This also affects beamforming codebooks.} We showed that interference-aware combining schemes based on the radiative near-field model can effectively exploit the extra degrees-of-freedom offered by the propagation channel to deal with interference so as to enhance the scalability (in terms of number of \acp{UE}) and fairness of the system. This applies already to 5G multi-user \ac{MIMO} communications above $6$\,GHz (e.g., in the range of mmWave bands), and will be further exacerbated by beyond-5G communications, operating in the sub-THz spectrum. %\textcolor{blue}{We thus hope to stimulate the interest of the research community towards the use of appropriate channel models to push further the limits of multi-user \ac{MIMO} communications.}

\vspace{-0.4cm}
\bibliographystyle{IEEEtran}
\bibliography{mybib}

\end{document}

%% file: acronymsList.tex
\begin{acronym}[DVB-S2X]
\acro{BS}{base station}
\acro{CDF}{cumulative distribution function}
\acro{CSI}{channel state information}
\acro{MAI}{multiple access interference}
\acro{MIMO}{multiple-input multiple-output}
\acro{LoS}{line-of-sight}
\acro{MMSE}{minimum mean-square-error}
\acro{MU}{multi-user}
\acro{MR}{maximum ratio}
\acro{SE}{spectral efficiency}
\acro{SINR}{signal-to-interference-plus-noise ratio}
\acro{SNR}{signal-to-noise ratio}
\acro{ULA}{uniform linear array}
\acro{UE}{user equipment}
\acro{THz}{terahertz}
\end{acronym}

%% file: main.bbl
\begin{thebibliography}{10}
\providecommand{\url}[1]{#1}
\csname url@samestyle\endcsname
\providecommand{\newblock}{\relax}
\providecommand{\bibinfo}[2]{#2}
\providecommand{\BIBentrySTDinterwordspacing}{\spaceskip=0pt\relax}
\providecommand{\BIBentryALTinterwordstretchfactor}{4}
\providecommand{\BIBentryALTinterwordspacing}{\spaceskip=\fontdimen2\font plus
\BIBentryALTinterwordstretchfactor\fontdimen3\font minus
  \fontdimen4\font\relax}
\providecommand{\BIBforeignlanguage}[2]{{%
\expandafter\ifx\csname l@#1\endcsname\relax
\typeout{** WARNING: IEEEtran.bst: No hyphenation pattern has been}%
\typeout{** loaded for the language `#1'. Using the pattern for}%
\typeout{** the default language instead.}%
\else
\language=\csname l@#1\endcsname
\fi
#2}}
\providecommand{\BIBdecl}{\relax}
\BIBdecl

\bibitem{RappaportAccess2019}
T.~S. Rappaport, Y.~Xing, O.~Kanhere, S.~Ju, A.~Madanayake, S.~Mandal,
  A.~Alkhateeb, and G.~C. Trichopoulos, ``Wireless communications and
  applications above {100 GHz}: Opportunities and challenges for {6G} and
  beyond,'' \emph{IEEE Access}, vol.~7, pp. 78\,729--78\,757,
  {Jun.} 2019.

\bibitem{rel17}
{3rd Generation Partnership Project (3GPP)}, ``Technical specification group
  radio access network; {NR}; {Physical} channels and modulation ({Release}
  17),'' Tech. Rep. 3GPP TS 38.211 V17.3.0, Sep. 2022.

\bibitem{heath_lozano_2018}
R.~W. Heath~Jr. and A.~Lozano, \emph{Foundations of {MIMO}
  Communication}.\hskip 1em plus 0.5em minus 0.4em\relax Cambridge University
  Press, 2018.

\bibitem{Selvan2017a}
K.~T. Selvan and R.~Janaswamy, ``{Fraunhofer} and {Fresnel} distances: Unified
  derivation for aperture antennas,'' \emph{IEEE Ant. Prop. Mag.}, vol.~59,
  no.~4, pp. 12--15, {Aug.} 2017.

\bibitem{LozanoMag2021}
H.~Do, S.~Cho, J.~Park, H.-J. Song, N.~Lee, and A.~Lozano, ``Terahertz
  line-of-sight {MIMO} communication: Theory and practical challenges,''
  \emph{IEEE Commun. Mag.}, vol.~59, no.~3, {Mar.} 2021.

\bibitem{bohagen09}
F.~B{\o}hagen, P.~Orten, and G.~E. {\O}ien, ``On spherical vs. plane wave
  modeling of line-of-sight {MIMO} channels,'' \emph{IEEE Trans. Commun.},
  vol.~57, no.~3, Mar. 2009.

\bibitem{Madhow2011}
E.~Torkildson, U.~Madhow, and M.~Rodwell, ``Indoor millimeter wave {MIMO}:
  Feasibility and performance,'' \emph{IEEE Trans. Wireless Commun.}, vol.~10,
  no.~12, pp. 4150--4160, {Dec.} 2011.

\bibitem{Friedlander2019}
B.~Friedlander, ``Localization of signals in the near-field of an antenna
  array,'' \emph{IEEE Trans. Signal Process.}, vol.~67, no.~15,
  {Aug.} 2019.

\bibitem{imt2020}
{Radiocommunication sector of International Telecommunication Union (ITU-R)},
  ``Minimum requirements related to technical performance for {IMT-2020} radio
  interface(s),'' Tech. Rep. ITU-R M.2410-0, Nov. 2017.

\bibitem{bjornson2020}
E.~Bj{\"o}rnson and L.~Sanguinetti, ``Power scaling laws and near-field
  behaviors of massive {MIMO} and intelligent reflecting surfaces,'' \emph{IEEE
  Open J. Commun. Society}, vol.~1, pp. 1306--1324, {Sep.}
  2020.

\bibitem{dardari2020}
D.~Dardari, ``Communicating with large intelligent surfaces: Fundamental limits
  and models,'' \emph{IEEE J. Sel. Areas Commun.}, vol.~38, no.~11, pp.
  2526--2537, {Nov.} 2020.

\bibitem{demir2021}
E.~Bj{\"o}rnson, {\"O}.~T. Demir, and L.~Sanguinetti, ``A primer on near-field
  beamforming for arrays and reconfigurable intelligent surfaces,'' in
  \emph{Proc. Asilomar}, Pacific Grove, CA, USA, Nov. 2021.

\bibitem{bjornson2017}
E.~Bj{\"o}rnson, J.~Hoydis, and L.~Sanguinetti, ``Massive {MIMO} networks:
  Spectral, energy, and hardware efficiency,'' \emph{Foundations and Trends in
  Signal Processing}, vol. 3-4, no.~11, 2017.

\bibitem{air6419}
\BIBentryALTinterwordspacing
Ericsson. (2022) {Massive} {MIMO} solutions accelerate {5G} mid-band. [Online].
  Available: \url{https://www.ericsson.com/en/ran/massive-mimo}
\BIBentrySTDinterwordspacing

\bibitem{lu2021}
H.~Lu and Y.~Zeng, ``How does performance scale with antenna number for
  extremely large-scale {MIMO}?'' in \emph{Proc. IEEE Int. Conf. Commun.
  (ICC)}, Montreal, Canada, Jun. 2021.

\end{thebibliography}
